\documentclass[reqno]{amsart}

\usepackage{booktabs} 
\usepackage{IEEEtrantools}
\usepackage{amssymb,latexsym,amsfonts,amsmath}
\usepackage{graphicx}
\usepackage{dsfont}

\usepackage{algorithm}
\usepackage{algorithmic}

\newtheorem{theorem}{Theorem}[section]
\newtheorem{lemma}[theorem]{Lemma}

\newtheorem{corollary}[theorem]{Corollary}

\newtheorem{definition}[theorem]{Definition}
\newtheorem{example}[theorem]{Example}

\newtheorem{remark}[theorem]{Remark}
\newtheorem{assumption}{Assumption}
\numberwithin{equation}{section}

\newcommand{\R}{{\mathbb{R}}}

\newcommand{\N}{{\mathbb{N}}}

\newcommand{\Lab}{\mathsf{L}}
\newcommand{\until}{\mathbin{\sf U}}

\newcommand{\nex}{\mathord{\bigcirc}}
\newcommand{\trans}{\mathsf{t}}
\newcommand{\AP}{AP} 
\newcommand{\alphabeth}{\Sigma_{\textsf{a}}}
\newcommand{\word}{\omega}

\newcommand{\Let}{:=}
\newcommand{\EE}{\mathds{E}}
\newcommand{\PP}{\mathds{P}}

\begin{document}

\begin{abstract}
This paper is concerned with a compositional approach for constructing infinite abstractions of interconnected discrete-time stochastic control systems. The proposed approach uses the interconnection matrix and joint dissipativity-type properties of subsystems and their abstractions described by a new notion of so-called stochastic storage functions. The interconnected abstraction framework is based on new notions of so-called stochastic simulation functions, constructed compositionally using stochastic storage functions of components. Using stochastic simulation functions, one can quantify the distance between original interconnected stochastic control systems and interconnected abstractions in the probabilistic setting. Accordingly, one can leverage the proposed results to perform analysis and synthesis over abstract interconnected systems, and then carry the results back over concrete ones. In the first part of the paper, we derive dissipativity-type compositional reasoning for the quantification of the distance in probability between the interconnection of stochastic control subsystems and that of their abstractions. Moreover, we focus on a class of discrete-time nonlinear stochastic control systems with independent noises in the abstract and concrete subsystems, and propose a computational scheme to construct abstractions together with their corresponding stochastic storage functions. In the second part of the paper,  we consider specifications expressed as syntactically co-safe linear temporal logic formulae and show how a synthesized policy for the abstract system can be refined to a policy for the original system while providing guarantee on the probability of satisfaction. We demonstrate the effectiveness of the proposed results by constructing an abstraction (totally 3 dimensions) of the interconnection of three discrete-time nonlinear stochastic control subsystems (together 222 dimensions) in a compositional fashion such that the compositionality condition does not require any constraint on the number or gains of the subsystems. We also employ the abstraction as a substitute to synthesize a controller enforcing a syntactically co-safe linear temporal logic specification. 
\end{abstract}

\title[Compositional Construction of Infinite Abstractions for Networks of Stochastic Systems]{Compositional Construction of Infinite Abstractions for Networks of Stochastic Control Systems}

\author{Abolfazl Lavaei$^1$}
\author{Sadegh Soudjani$^2$}
\author{Majid Zamani$^{3,4}$}
\address{$^1$Department of Electrical and Computer Engineering, Technical University of Munich, Germany.}
\email{lavaei@tum.de}
\address{$^2$School of Computing, Newcastle University, United Kingdom}
\email{sadegh.soudjani@newcastle.ac.uk}
\address{$^3$Department of Computer Science, University of Colorado Boulder, USA}
\address{$^4$Department of Computer Science, Ludwig Maximilian University of Munich, Germany}
\email{majid.zamani@colorado.edu}
\maketitle

\section{Introduction}
Large-scale interconnected systems have received significant attentions in the last few years due to their presence in real life systems including power networks and air traffic control. Each complex real-world system can be regarded as an interconnected system composed of several subsystems. Since these large-scale network of systems are inherently difficult to analyze and control, one can develop compositional schemes and employ the abstractions of the given networks as a replacement in the controller design process. In other words, in order to overcome the computational complexity in large-scale interconnected systems, one can abstract the original concrete system by a simpler one with potentially a lower dimension. Those abstractions allow us to design controllers for them, and then refine the controllers to the ones for the concrete complex systems, while provide us with the quantified errors in this controller synthesis detour.

In the past few years, there have been several results on the construction of (in)finite abstractions for stochastic systems.
Existing results for \emph{continuous-time} systems include infinite approximation techniques for jump-diffusion systems \cite{julius2009approximations}, finite bisimilar abstractions for incrementally stable stochastic switched systems \cite{zamani2015symbolic} and randomly switched stochastic systems \cite{zamani2014approximately}, and finite bisimilar abstractions for incrementally stable stochastic control systems without discrete dynamics \cite{zamani2014symbolic}. Recently, compositional construction of infinite abstractions is discussed in \cite{zamani2016approximations} using small-gain type conditions and of finite bisimilar abstractions in \cite{2017arXiv170909546M} based on a new notion of disturbance bisimilarity relation.

For \emph{discrete-time} stochastic models with continuous state spaces, finite abstractions are initially employed in \cite{APLS08} for formal synthesis of this class of systems. The algorithms are improved in terms of scalability in \cite{SA13,SSoudjani} and implemented in the tool \texttt{FAUST} \cite{FAUST15}. Extension of the techniques to infinite horizon properties is proposed in \cite{tkachev2011infinite} and formal abstraction-based policy synthesis is discussed in \cite{tmka2013}. A new notion of approximate similarity relation is proposed in \cite{HS19,SSA16} that takes into account both deviation in stochastic evolution and in outputs of the two systems. Compositional construction of infinite abstractions (reduced order models) using small-gain type conditions is proposed in \cite{lavaei2017compositional} . Compositional construction of finite abstractions is discussed in \cite{SAM15},~\cite{lavaei2017HSCC}, and~\cite{lavaei2018ADHS} using dynamic Bayesian networks, dissipativity-type reasoning, and small-gain conditions, respectively, all for discrete-time stochastic control systems. Recently, compositional synthesis of large-scale stochastic systems using a relaxed dissipativity approach is proposed in~\cite{lavaei2019NAHS}. Compositional (in)finite abstractions for large-scale interconnected stochastic systems using small-gain type conditions are proposed in~\cite{lavaei2018ADHSJ}.

In this paper, we provide a compositional approach for the construction of infinite abstractions of interconnected discrete-time stochastic control systems using the interconnection matrix and joint dissipativity-type properties of subsystems and their abstractions. Our abstraction framework is based on a new notion of stochastic simulation functions under which an abstraction, which is itself a discrete-time stochastic control system with potentially a lower dimension, performs as a substitute in the controller design process. The stochastic simulation function is used to quantify the error in probability in this detour controller synthesis scheme. As a consequence, one can leverage our proposed results to synthesize a policy that satisfies a temporal logic property over the abstract interconnected system and then refine this policy back for the concrete interconnected one.

Our proposed approach differs from the one in \cite{lavaei2017compositional} in three directions. First and foremost, rather than using small-gain type reasoning, we employ the dissipativity-type compositional reasoning that may not require any constraint on the number or gains of the subsystems for some interconnection topologies (cf. case study). Second, we provide a scheme for the construction of infinite abstractions for a class of discrete-time nonlinear stochastic control systems whereas the construction scheme in \cite{lavaei2017compositional} only handles linear systems. As our third contribution, we consider a fragment of linear temporal logic (LTL) known as syntactically co-safe linear temporal logic (scLTL) \cite{KupfermanVardi2001} whereas the results in  \cite{lavaei2017compositional} only deal with finite-horizon invariant.
In particular, given such a specification over the concrete system, we construct an epsilon-perturbed specification over the abstract system whose probability of satisfaction gives a lower bound for the probability of satisfaction in the concrete domain. 

It should be also noted that we do not put any restriction on the sources of uncertainties in the concrete and abstract systems. Thus our result is more general than \cite{zamani2016approximations}, where the noises in the concrete and abstract systems are assumed to be the same, which means the abstraction has access to the noise of the concrete system. Finally, we show the effectiveness of dissipativity-type compositional reasoning for large-scale systems by first constructing an abstraction (totally 3 dimensions) of the interconnection of three discrete-time nonlinear stochastic control subsystems (together 222 dimensions) in a compositional fashion. Then, we employ the abstraction as a substitute to synthesize a controller enforcing a syntactically co-safe linear temporal logic specification over the concrete network.

\section{Discrete-Time Stochastic Control Systems}\label{Sec: dt-SCS}
\subsection{Preliminaries}
We consider a probability space $(\Omega,\mathcal F_{\Omega},\PP_{\Omega})$,
where $\Omega$ is the sample space,
$\mathcal F_{\Omega}$ is a sigma-algebra on $\Omega$ comprising subsets of $\Omega$ as events, and $\PP_{\Omega}$ is a probability measure that assigns probabilities to events. We assume that random variables introduced in this article are measurable functions of the form $X:(\Omega,\mathcal F_{\Omega})\rightarrow (S_X,\mathcal F_X)$.
Any random variable $X$ induces a probability measure on  its space $(S_X,\mathcal F_X)$ as $Prob\{A\} = \PP_{\Omega}\{X^{-1}(A)\}$ for any $A\in \mathcal F_X$.
We often directly discuss the probability measure on $(S_X,\mathcal F_X)$ without explicitly mentioning the underlying probability space and the function $X$ itself.

A topological space $S$ is called a Borel space if it is homeomorphic to a Borel subset of a Polish space (i.e., a separable and completely metrizable space).
Examples of a Borel space are the Euclidean spaces $\mathbb R^n$, its Borel subsets endowed with a subspace topology, as well as hybrid spaces.
Any Borel space $S$ is assumed to be endowed with a Borel sigma-algebra, which is
denoted by $\mathcal B(S)$. We say that a map $f : S\rightarrow Y$ is measurable whenever it is Borel measurable.

\subsection{Notation}
The following notation is used throughout the paper. We denote the set of nonnegative integers by $\mathbb N := \{0,1,2,\ldots\}$ and the set of positive integers by $\mathbb N_{\ge 1} := \{1,2,3,\ldots\}$. The symbols $\R$, $\R_{>0}$, and $\R_{\ge 0}$ denote the set of real, positive and nonnegative real numbers, respectively. Given a vector $x\in\mathbb{R}^{n}$, $\Vert x\Vert$ denotes the Euclidean norm of $x$.
Symbols $I_n$ and $\mathbf{1}_n$ denote respectively the identity matrix in $\R^{n\times{n}}$ and the column vector in $\R^{n\times{1}}$ with all elements equal to one. Given $N$ vectors $x_i \in \R^{n_i}$, $n_i\in \mathbb N_{\ge 1}$, and $i\in\{1,\ldots,N\}$, we use $x = [x_1;\ldots;x_N]$ to denote the corresponding vector of dimension $\sum_i n_i$. We denote by $\mathsf{diag}(a_1,\ldots,a_N)$ a diagonal matrix in $\R^{N\times{N}}$ with diagonal matrix entries $a_1,\ldots,a_N$ starting from the upper left corner. Given functions $f_i:X_i\rightarrow Y_i$, for any $i\in\mathbb\{1,\ldots,N\}$, their Cartesian product $\prod_{i=1}^{N}f_i:\prod_{i=1}^{N}X_i\rightarrow\prod_{i=1}^{N}Y_i$ is defined as $(\prod_{i=1}^{N}f_i)(x_1,\ldots,x_N)=[f_1(x_1);\ldots;f_N(x_N)]$.
For any set $A$ we denote by $A^{\mathbb N}$ the Cartesian product of a countable number of copies of $A$, i.e., $A^{\mathbb N} = \prod_{k=0}^{\infty} A$. A function $\gamma:\R_{\ge 0}\rightarrow\R_{\ge 0}$, is said to be a class $\mathcal{K}$ function if it is continuous, strictly increasing and $\gamma(0)=0$. A class $\mathcal{K}$ function $\gamma$ is said to be a class $\mathcal{K}_{\infty}$ if $\gamma(r)\rightarrow\infty$ as $r\rightarrow\infty$.

\subsection{Discrete-Time Stochastic Control Systems}

We consider stochastic control systems in discrete time (dt-SCS) defined over a general state space and characterized by the tuple
\begin{equation}
	\label{eq:dt-SCS}
	\Sigma=\left(X,U,W,\varsigma,f,Y_1,Y_2, h_1, h_2\right)\!,
\end{equation}
where $X$ is a Borel space as the state space of the system.
We denote by $(X, \mathcal B (X))$ the measurable space
with $\mathcal B (X)$  being  the Borel sigma-algebra on the state space. Sets
$U$ and $W$ are Borel spaces as the \emph{external} and \emph{internal} input spaces of the system. Notation $\varsigma$ denotes a sequence of independent and identically distributed (i.i.d.) random variables on a set $V_\varsigma$ as
\begin{equation*}
	\varsigma:=\{\varsigma(k):\Omega\rightarrow V_{\varsigma},\,\,k\in\N\}.
\end{equation*}
The map $f:X\times U\times W\times V_{\varsigma} \rightarrow X$ is a measurable function characterizing the state evolution of the system.	
Finally, sets $Y_1$ and $Y_2$ are Borel spaces as the external and internal output spaces of the system, respectively.
Maps $h_1:X\rightarrow Y_1$ and $h_2:X\rightarrow Y_2$ are measurable functions that map a state $x\in X$ to its external and internal outputs $y_1 = h_1(x)$ and $y_2 = h_2(x)$, respectively.

For given initial state $x(0)\in X$ and input sequences $\nu(\cdot):\mathbb N\rightarrow U$ and $w(\cdot):\mathbb N\rightarrow W$, evolution of the state of dt-SCS $\Sigma$ can be written as
\begin{equation}\label{Eq_1a}
	\Sigma:\left\{\hspace{-1mm}\begin{array}{l}x(k+1)=f(x(k),\nu(k),w(k),\varsigma(k)),\\
		y_1(k)=h_1(x(k)),\\
		y_2(k)=h_2(x(k)),\\
	\end{array}\right.
	\quad k\in\mathbb N.
\end{equation}

\begin{remark}
	The above definition can be generalized by allowing the set of valid external inputs to depend on the current state and internal input of the system, i.e.,
	to include $\{U(x,w)|x\in X,w\in W\}$ in the definition of dt-SCS which is a family of non-empty measurable subsets of $U$ with the property that
	\begin{equation}\notag
		K :=\{(x,\nu, w): x\in X,\,w\in W,\,\nu\in U(x,w)\},
	\end{equation}
	is measurable in $X\times U\times W$.
	For the succinct presentation of the results, we assume in this paper that the set of valid external inputs is the whole external input space: $U(x,w) = U$ for all $x\in X$ and $w\in W$, but the obtained results are generally applicable.
\end{remark}

Given the dt-SCS in \eqref{eq:dt-SCS}, we are interested in \emph{Markov policies} to control the system.
\begin{definition}
	A Markov policy for the dt-SCS $\Sigma$ in \eqref{eq:dt-SCS} is a sequence
	$\gamma = (\gamma_0,\gamma_1,\gamma_2,\ldots)$ of universally measurable stochastic kernels $\gamma_n$ \cite{BS96},
	each defined on the input space $U$ given $X\times W$ and such that for all $(x_n,w_n)\in X\times W$, $\gamma_n(U|(x_n,w_n))=1$.
	The class of all such Markov policies is denoted by $\Pi_M$. 
\end{definition} 

We associate respectively to $U$ and $W$ the sets $\mathcal U$ and $\mathcal W$ to be collections of sequences $\{\nu(k):\Omega\rightarrow U,\,\,k\in\N\}$ and $\{w(k):\Omega\rightarrow W,\,\,k\in\N\}$, in which $\nu(k)$ and $w(k)$ are independent of $\varsigma(t)$ for any $k,t\in\mathbb N$ and $t\ge k$. For any initial state $a\in X$, $\nu(\cdot)\in\mathcal{U}$, and $w(\cdot)\in\mathcal{W}$,
the random sequences $x_{a\nu w}:\Omega \times\N \rightarrow X$, $y^1_{a\nu w}:\Omega \times \N \rightarrow Y_1$ and $y^2_{a\nu w}:\Omega \times \N \rightarrow Y_2$ that satisfy \eqref{Eq_1a}
are called respectively the \textit{solution process} and external and internal \textit{output trajectory} of $\Sigma$ under external input $\nu$, internal input $w$ and initial state $a$.

\begin{remark}
	In this paper, we are ultimately interested in investigating discrete-time stochastic control systems without internal inputs and outputs. In this case, the tuple~\eqref{eq:dt-SCS} reduces to $(X,U,\varsigma,f,Y,h)$ and dt-SCS~\eqref{Eq_1a} can be re-written as
	\begin{equation}\notag
		\Sigma:\left\{\hspace{-1mm}\begin{array}{l}x(k+1)=f(x(k),\nu(k),\varsigma(k)),\\
			y(k)=h(x(k)),\\
		\end{array}\right.
		\quad k\in\mathbb N.
	\end{equation}	  	
	The interconnected control systems, defined later, are also a class of control systems without internal signals, resulting from the interconnection of dt-SCSs having both internal and external inputs and outputs.  
\end{remark}

In the sequel we assume that the state and output spaces $X$ and $Y$ of $\Sigma$ are subsets of $\mathbb R^n$ and $\mathbb R^q$, respectively. System $\Sigma$ is called finite if $ X, U, W$ are finite sets and infinite otherwise.

\section{Stochastic Storage and Simulation Functions}\label{Sec: Simulation Funcation}
In this section, we first introduce a notion of so-called stochastic storage functions for the discrete-time stochastic control systems with both internal and external inputs which is adapted from the notion of storage functions from
dissipativity theory \cite{2016Murat}. We then define a notion of stochastic
simulation functions for systems with only external input. We use these definitions to quantify closeness of two dt-SCS.

\begin{definition}\label{Def_1a}
	Consider dt-SCS $\Sigma =(X,U,W, \varsigma, f,Y_1,Y_2,h_1,h_2)$ and $\widehat\Sigma =(\hat X,\hat U, \hat W, \hat \varsigma, \hat f, Y_1, \hat Y_2, \hat h_1, \hat h_2)$ with the same external output spaces. A function $V:X\times\hat X\to\R_{\ge0}$ is called a stochastic storage function (SStF) from  $\widehat\Sigma$ to $\Sigma$ if there exist $\alpha\in\mathcal{K}_\infty$,~$\kappa\in \mathcal{K}$, $\rho_{\mathrm{ext}}\in\mathcal{K}_\infty\cup\{0\}$, some matrices $G,\hat G,H$ of appropriate dimensions, and some symmetric matrix $ \bar X$ of appropriate dimension with conformal block partitions $\bar X^{ij}$, $i,j\in\{1,2\}$, such that for any $x\in X$ and $\hat x\in\hat X$ one has
	\begin{align}\label{Eq_2a}
		\alpha(\Vert h_1(x)-\hat h_1(\hat x)\Vert)\le V(x,\hat x),
	\end{align}
	and $\forall x\in X$ $\forall\hat x\in\hat X$ $\forall\hat\nu\in\hat U$  $\exists \nu\in U$ such that $\forall \hat{w}\in\hat W$ $\forall w\in W$ one obtains
	\begin{align}\notag\label{Eq_3a}
		\EE &\Big[V(x(k+1),\hat{x}(k+1))\,\big|\,x(k) \!=\! x,\hat{x}(k) \!=\! \hat x, w(k)\!=\!w,\hat{w}(k)\!=\!\hat{w},\nu(k)\!=\!\nu,\hat{\nu}(k)\!=\!\hat{\nu}\Big]-V(x,\hat{x})\\ 
		&\leq\!-\kappa(V(x,\hat{x}))\!+\!\begin{bmatrix}
			Gw-\hat G\hat w\\
			h_2(x)-H\hat h_2(\hat x)
		\end{bmatrix}^T\overbrace{\begin{bmatrix}
				\bar X^{11}&\bar X^{12}\\
				\bar X^{21}&\bar X^{22}
		\end{bmatrix}}^{\bar X:=}\begin{bmatrix}
			Gw-\hat G\hat w\\
			h_2(x)-H\hat h_2(\hat x)
		\end{bmatrix}\!+\! 
		\rho_{\mathrm{ext}}(\Vert\hat\nu\Vert)\!+\!\psi,
	\end{align}
	for some $\psi \in\R_{\ge 0}$.
\end{definition}

We use notation $\widehat\Sigma\preceq_{\mathcal{S}}\Sigma$ if there exists a storage function $V$ from $\widehat\Sigma$ to $\Sigma$, in which $\widehat\Sigma$ is  considered as an abstraction of concrete system $\Sigma$.

\begin{remark}
	The second condition above implies implicitly existence of a function $\nu=\nu_{\hat \nu}(x,\hat x,\hat \nu)$ for satisfaction of~\eqref{Eq_3a}. This function is called the \emph{interface function} and can be used to refine a synthesized policy $\hat\nu$ for $\widehat\Sigma$ to a policy $\nu$ for $\Sigma$.
\end{remark}

For the dt-SCS without internal signals (including interconnected dt-SCS), the above notion reduces to the following definition.

\begin{definition}
	Consider two dt-SCS
	$\Sigma\!=\!(\!X,U,\varsigma, \!f, \!Y,\!h)$ and
	$\widehat\Sigma =(\hat X,\hat U, \hat \varsigma, \hat f, Y,\hat h)$
	with the same output spaces.
	A function $V:X\times\hat X\to\R_{\ge0}$ is
	called a \emph{stochastic simulation function} (SSF) from $\widehat\Sigma$  to $\Sigma$ if
	\begin{itemize}
		\item $\exists \alpha\in \mathcal{K}_{\infty}$ such that
		\begin{eqnarray}
			\label{eq:lowerbound2}
			\forall x\in X,\forall \hat x\in\hat X,\quad \alpha(\Vert h(x)-\hat h(\hat x)\Vert)\le V(x,\hat x),
		\end{eqnarray}
		\item $\forall x\in X,\hat x\in\hat X,\hat\nu\in\hat U$, $\exists \nu\in U$ such that
		\begin{align}
			\EE \Big[V(x(k\!+\!1),\hat{x}(k\!+\!1))\,\big|\,x(k) \!=\! x,\hat{x}(k) \!=\! \hat x,\nu(k)\!=\!\nu,\hat{\nu}(k)\!=\!\hat{\nu}\Big]\!-\!V(x,\hat{x})\!\leq\!-\kappa(V(x,\hat{x}))
			\!+\!\rho_{\mathrm{ext}}(\Vert\hat\nu\Vert)\!+\!\psi,\label{eq:martingale2}
		\end{align}
		for some $\kappa\in \mathcal{K}$, $\rho_{\mathrm{ext}} \in \mathcal{K}_{\infty}\cup \{0\}$, and $\psi \in\R_{\ge 0}$.
	\end{itemize}
\end{definition}

\begin{remark}
	Conditions~\eqref{Eq_3a},\eqref{eq:martingale2} roughly speaking guarantee that if the concrete system and its abstraction start from two close initial conditions, then they remain close (in terms of expectation) after one step (i.e., roughly, if they start close, they will remain close). These types of conditions are closely related to the ones in the notions of (bi)simulation relations~\cite{tabuada2009verification}.
\end{remark}
In order to show the usefulness of SSF in comparing output trajectories of two dt-SCS in a probabilistic setting, we need the following technical lemma borrowed from~\cite[Theorem 3, pp. 86]{1967stochastic} with some slight modifications for the finite-time horizon, and also~\cite[Theorem 12, pp. 71]{1967stochastic} for the infinite-time horizon.

\begin{lemma}\label{Lemma: Kushner}
	Let $\Sigma =(X, \varsigma, f, Y,h)$ be a dt-SCS with the transition map $f:X\times V_\varsigma\rightarrow X$.\\
	i) \textsf{Finite-time horizon}: Assume there exist $V:X\to\R_{\ge0}$ and constants $0<\widehat\kappa<1$, and $\widehat \psi \in\R_{\ge 0}$ such that
	\begin{align}\notag
		&\EE \Big[V(x(k+1))\big|x(k) = x\Big]\leq\widehat\kappa V(x) + \widehat\psi.
	\end{align}
	Then for any random variable $a$ as the initial state of the dt-SCS, the following inequity holds:
\begin{equation}\notag
\PP\left\{\sup_{0\leq k\leq T_d} V(x)\geq\varepsilon\,|\,a\right\}\le\delta,
\end{equation}
\begin{equation*}
\delta := 
\begin{cases}
1-(1-\frac{V(a)}{\varepsilon})(1-\frac{\widehat\psi}{\varepsilon})^{T_d} & \quad \text{if}~\varepsilon\geq\frac{\widehat\psi}{\widehat\kappa},\\
(\frac{V(a)}{\varepsilon})(1-\widehat\kappa)^{T_d}+(\frac{\widehat\psi}{\widehat\kappa\varepsilon})(1-(1-\widehat\kappa)^{T_d}) & \quad\text{if}~\varepsilon<\frac{\widehat\psi}{\widehat\kappa}.
\end{cases}
\end{equation*}
	ii) \textsf{Infinite-time horizon}: Assume there exists a nonnegative $V:X\to\R_{\ge0}$ such that
	\begin{align}\notag
		&\EE \Big[V(x(k+1))\big|x(k) = x\Big]- V(x)\leq0.
	\end{align}
	Function $V$ satisfying the above inequality is also called supermartingale. Then for any random variable $a$ as the initial state of the dt-SCS, the following inequity holds:
	\begin{equation}\notag
		\PP\left\{\sup_{0\leq k< \infty} V(x)\geq\varepsilon\,|\,a\right\}\le\frac{V(a)}{\varepsilon}.
	\end{equation}
\end{lemma}

Now by employing Lemma~\ref{Lemma: Kushner}, we provide one of the results of the paper.

\begin{theorem}\label{Thm_1a}
	Let
	$\Sigma =(X,U, \varsigma, f, Y,h)$ and
	$\widehat\Sigma =(\hat X,\hat U, \hat \varsigma, \hat f, Y,\hat h)$
	be two dt-SCS
	with the same output spaces. Suppose $V$ is an SSF from $\widehat\Sigma$ to $\Sigma$, and there exists a constant $0<\widehat\kappa<1$ such that function $\kappa \in \mathcal{K}$ in~\eqref{eq:martingale2} satisfies $\kappa(r)\geq\widehat\kappa r$ $\forall r\in\R_{\geq0}$. For any random variables $a$ and $\hat{a}$ as the initial states of the two dt-SCS and any external input trajectory $\hat\nu(\cdot)\in\mathcal{\hat U}$ preserving Markov property for the closed-loop $\widehat \Sigma$,
	there exists an input trajectory $\nu(\cdot)\in\mathcal{U}$ of $\Sigma$ through the interface function associated with $V$ such that the following inequality holds
	\begin{equation}\notag
	\PP\left\{\sup_{0\leq k\leq T_d}\Vert y_{a\nu}(k)-\hat y_{\hat a \hat\nu}(k)\Vert\geq\varepsilon\,|\,[a;\hat a]\right\}\le\delta,
	\end{equation}
	\begin{equation}\label{Eq_25}
	\delta := 
	\begin{cases}
	1-(1-\frac{V(a,\hat a)}{\alpha\left(\varepsilon\right)})(1-\frac{\widehat\psi}{\alpha\left(\varepsilon\right)})^{T_d} & \quad \text{if}~\alpha\left(\varepsilon\right)\geq\frac{\widehat\psi}{\widehat\kappa},\\
	(\frac{V(a,\hat a)}{\alpha\left(\varepsilon\right)})(1-\widehat\kappa)^{T_d}+(\frac{\widehat\psi}{\widehat\kappa\alpha\left(\varepsilon\right)})(1-(1-\widehat\kappa)^{T_d}) & \quad\text{if}~\alpha\left(\varepsilon\right)<\frac{\widehat\psi}{\widehat\kappa},
	\end{cases}
	\end{equation}
	provided that there exists a constant $\widehat\psi\geq0$ satisfying  $\widehat\psi\geq \rho_{\mathrm{ext}}(\Vert\hat \nu\Vert_{\infty})+\psi$.
\end{theorem}
The proof of Theorem~\ref{Thm_1a} is provided in the Appendix. The results shown in Theorem~\ref{Thm_1a} provide closeness of output behaviours of two systems in finite-time horizon. We can extend the result to infinite-time horizon using Lemma~\ref{Lemma: Kushner} given that $\widehat{\psi}=0$ as stated in the following corollary.

\begin{corollary}\label{corollary: Supermartingle}
	Let $\Sigma$ and $\widehat\Sigma$ be two dt-SCS without internal inputs and outputs and with the same output spaces.
	Suppose $V$ is an SSF from $\widehat\Sigma$ to $\Sigma$ such that $\rho_{\mathrm{ext}}(\cdot)\equiv0$ and $\psi = 0$.
	For any random variables $a$ and $\hat{a}$ as the initial states of the two dt-SCS and any external input trajectory $\hat\nu(\cdot)\in\mathcal{\hat U}$ preserving Markov property for the closed-loop $\widehat \Sigma$, there exists $\nu(\cdot)\in{\mathcal{U}}$ of $\Sigma$ through the interface function associated with $V$ such that the following inequality holds:
	\begin{align}\nonumber
		\PP\left\{\sup_{0\leq k< \infty}\Vert y_{a\nu}(k)-\hat y_{\hat a 0}(k)\Vert\ge \varepsilon\,|\,[a;\hat a]\right\}\leq\frac{V(a,\hat a)}{\alpha\left(\varepsilon\right)}.
	\end{align}
\end{corollary}
The proof of Corollary~\ref{corollary: Supermartingle} is provided in the Appendix.
\begin{remark}
	Note that $\psi=0$ is possible mainly if concrete and abstract systems are both continuous-space but possibly with different dimensions and share the same multiplicative noise. Depending on the dynamic, function $\rho_{\mathrm{ext}}(\cdot)$ can be identically zero (cf. Section \ref{Sec: Nonlinear Control} and case study).
\end{remark}

The relation \eqref{Eq_25} lower bounds the probability such that the Euclidean distance between any output trajectory of the abstract model and the corresponding one of the concrete model remains close and is different from the probabilistic version discussed for finite state, discrete-time labeled Markov chains in \cite{DLT08}, which hinges on the absolute difference between transition probabilities over sets covering the state space. However, one can still use the results in Theorem \ref{Thm_1a} and design controllers for abstractions and refine them to the ones for concrete systems while providing the probability of satisfaction over the concrete domain which is discussed in detail later in Section~\ref{Sec: Finite Automata}.
\section{Compositional Abstractions for Interconnected Systems}
\label{Sec: Compositionality}

In this section, we analyze networks of control systems and show how to construct their abstractions together with the corresponding simulation functions by using stochastic storage functions for the subsystems. We first provide a formal definition of interconnection between discrete-time stochastic control subsystems.
\begin{definition}
	Consider $N\in\N_{\geq1}$ stochastic control subsystems $\Sigma_i=(X_i,U_i,W_i,\varsigma_i, f_i, Y_{1i},Y_{2i},h_{1i},h_{2i})$, $i\in\{1,\ldots,N\}$, and a static matrix $M$ of an appropriate dimension defining the coupling of these subsystems. The
	interconnection of  $\Sigma_i$ for any $i\in\{1,\ldots,N\}$, is the interconnected stochastic control system $\Sigma=(X,U,\varsigma, f, Y,h)$, denoted by
	$\mathcal{I}(\Sigma_1,\ldots,\Sigma_N)$, such that $X:=\prod_{i=1}^{N}X_i$,  $U:=\prod_{i=1}^{N}U_i$, function $f:=\prod_{i=1}^{N}f_{i}$, $Y:=\prod_{i=1}^{N}Y_{1i}$, and $h=\prod_{i=1}^{N}h_{1i}$, with the
	internal variables constrained by:
	\begin{align}\notag
		[{w_{1};\ldots;w_{N}}]=M[{h_{21}(x_1);\ldots;h_{2N}(x_N)}].
	\end{align}
\end{definition}

\subsection{Compositional Abstractions of Interconnected Systems}
This subsection contains one of the main contributions of the paper. Assume that we are given $N$ stochastic control subsystems $\Sigma_i\!=\!(X_i,U_i, W_i,\varsigma_i, f_i, Y_{1i},\\,Y_{2i},h_{1i},h_{2i})$ together with their corresponding abstractions $\widehat\Sigma_i=(\hat X_i,\hat U_i,\hat W_i, \hat \varsigma_i, \hat f_i, Y_{1i},\hat Y_{2i},\hat h_{1i}, \hat h_{2i})$  with SStF $V_i$ from $\widehat\Sigma_i$ to $\Sigma_i$. We use $\alpha_{i}$, $\kappa_i$, $\rho_{\mathrm{ext}i}$, $H_i$, $G_i$, $\hat G_i$, $\bar X_i$, $\bar X_i^{11}$, $\bar X_i^{12}$, $\bar X_i^{21}$, and $\bar X_i^{22}$ to denote the corresponding functions, matrices, and their corresponding conformal block partitions appearing in Definition~\ref{Def_1a}.

In the next theorem, as one of the main results of the paper, we quantify the error between the interconnection of stochastic control subsystems and that of their abstractions in a compositional way.
\begin{theorem}\label{Thm_2a}
	Consider interconnected stochastic control system
	$\Sigma=\mathcal{I}(\Sigma_1,\ldots,\Sigma_N)$ induced by $N\in\N_{\geq1}$ stochastic
	control subsystems~$\Sigma_i$ and the coupling matrix $M$. Suppose stochastic control subsystems $\widehat \Sigma_i$ are abstractions of $\Sigma_i$ with the corresponding SStF $V_i$. If there exist $\mu_{i}>0$, $i\in\{1,\ldots,N\}$, and matrix $\hat M$ of appropriate dimension such that the matrix (in)equality
	\begin{align}
		\begin{bmatrix}\label{Con_1a}
			GM\\I_{\tilde q}
		\end{bmatrix}^T \!\!\!&\bar X_{cmp}\!\begin{bmatrix}
			GM\\I_{\tilde q}
		\end{bmatrix}\preceq0,
		\\\label{Con_2a}
		&\!\!\!\!\!\!GMH=\hat G\hat M,
	\end{align}
	are satisfied, where $\tilde q=\sum_{i=1}^Nq_{2i}$ and $q_{2i}$ are dimensions of internal outputs of subsystem $\Sigma_i$, and
	\begin{align}\notag
		&G:=\mathsf{diag}(G_1,\ldots,G_N),~\hat G:=\mathsf{diag}(\hat G_1,\ldots,\hat G_N),~H:=\mathsf{diag}(H_1,\ldots,H_N),\\
		&\bar X_{cmp}:=\begin{bmatrix}
			\mu_1\bar X_1^{11}&&&\mu_1\bar X_1^{12}&&\\
			&\ddots&&&\ddots&\\
			&&\mu_N\bar X_N^{11}&&&\mu_N\bar X_N^{12}\\
			\mu_1\bar X_1^{21}&&&\mu_1\bar X_1^{22}&&\\
			&\ddots&&&\ddots&\\
			&&\mu_N\bar X_N^{21}&&&\mu_N\bar X_N^{22}
		\end{bmatrix}\!\!,\label{Def_3a}
	\end{align}
	then
	\begin{align}\label{eq565}
		V(x,\hat x)\Let\sum_{i=1}^N\mu_iV_i(x_i,\hat x_i),
	\end{align}
	is a stochastic simulation function from the interconnected control system $\widehat \Sigma=\mathcal{I}(\hat
	\Sigma_1,\ldots,\widehat \Sigma_N)$, with the coupling matrix $\hat M$, to $\Sigma$.
\end{theorem}
The proof of Theorem~\ref{Thm_2a} is provided in the Appendix.
Note that matrix $\bar X_{cmp}$ in~\eqref{Def_3a} has zero matrices in all its empty entries.

\begin{remark}
	Linear matrix inequality (LMI) \eqref{Con_1a} with $G = I$ is similar to the LMI appearing in \cite[Chapter 2]{2016Murat} for a compositional stability condition based on dissipativity theory. As discussed in  \cite{2016Murat}, the LMI holds independently of the number of subsystems in many physical applications with specific interconnection structures including communication networks, flexible joint robots, power generators, and so on. We refer the interested readers to \cite{2016Murat} for more details on the satisfaction of this type of LMI.
\end{remark}
\begin{remark}
	One can relax condition~\eqref{Con_2a} and employ the linear least square approach instead of solving the equality exactly. In this case, an additional error resulting from the least square approach is added to $\psi$ in~\eqref{Eq_4a} which is left for the future investigations.
\end{remark}

\section{Discrete-Time Stochastic Control Systems with Slope Restrictions on Nonlinearity}\label{Sec: Nonlinear Control}
In this section, we focus on a specific class of discrete-time nonlinear stochastic control systems $\Sigma_{\textsf{nl}}$ and \emph{quadratic} stochastic storage functions $V$ and provide an approach on the construction of their abstractions. In the next subsection, we first formally define the class of discrete-time nonlinear stochastic control systems.

\subsection{A Class of Discrete-Time Nonlinear Stochastic Control Systems}\label{Subsec: Nonlinear Control}

The class of discrete-time nonlinear stochastic control systems, considered here, is given by
\begin{align}\label{Eq_5a}
	\Sigma_{\textsf{nl}}:\left\{\hspace{-1.5mm}\begin{array}{l}x(k+1)=Ax(k)+E\varphi(Fx(k))+B\nu(k)+Dw(k)+R\varsigma(k),\\
		y_1(k)=C_1x(k),\\
		y_2(k)=C_2x(k),\end{array}\right.
\end{align}
where the additive noise $\varsigma(k)$ is a sequence of independent random vectors with multivariate standard normal distributions, and $\varphi:\R\rightarrow\R$ satisfies the following constraint
\begin{equation}\label{Eq_6a}
	0\leq\frac{\varphi(v)-\varphi(w)}{v-w}\leq b,~~~\forall v,w\in\R,v\neq w,
\end{equation}
for some $b\in\R_{>0}\cup\{\infty\}$.

We use the tuple
\begin{align}\notag
	\Sigma_{\textsf{nl}}=(A,B,C_1,C_2,D,E,F,R,\varphi),
\end{align}
to refer to the class of discrete-time nonlinear stochastic control systems of the form~\eqref{Eq_5a}.

If $\varphi$ in~\eqref{Eq_5a} is linear including the zero function (i.e. $\varphi\equiv0$) or $E$ is a zero matrix, one can remove or push the term $E\varphi(Fx)$ to $Ax$ and, hence, the tuple representing the class of discrete-time nonlinear stochastic control systems reduces to the linear one $\Sigma_{\textsf{nl}}=(A,B,C_1,C_2,D, R)$. Therefore, every time we use the tuple $\Sigma_{\textsf{nl}}=(A,B,C_1,C_2,D,E,F,R,\varphi)$, it implicitly implies that $\varphi$ is nonlinear and $E$ is nonzero.

\begin{remark}
	Although the lower bound in~\eqref{Eq_6a} is zero, one can also assume \eqref{Eq_6a} with some nonlinear functions $\varphi$ with a nonzero lower-bound, e.g., $a\in\R$. In this case, one can make a change of coordinate and define a new function $\tilde\varphi(r):=\varphi(r)-ar$ which satisfies~\eqref{Eq_6a} with $\tilde a=0$ and $\tilde b=b-a$, and rewrite~\eqref{Eq_5a} as
	\begin{align}\notag
		\Sigma_{\textsf{nl}}:\left\{\hspace{-1.5mm}\begin{array}{l}x(k+1)=\tilde Ax(k)+E\tilde \varphi(Fx(k))+B\nu(k)+Dw(k)+R\varsigma(k),\\
			y_1(k)=C_1x(k),\\
			y_2(k)=C_2x(k),\end{array}\right.
	\end{align}
	where $\tilde A=A+aEF$.
\end{remark}

In the next subsection, we provide conditions under which a candidate $V$ is an SStF facilitating the construction of an abstraction $\widehat \Sigma_{\textsf{nl}}$.
\subsection{Quadratic Stochastic Storage Functions}\label{Subsec: Quadratic Stochastic Func}
Here, we employ the following quadratic SStF
\begin{align}\label{Eq_7a}
	V(x,\hat x)=(x-P\hat x)^T\tilde M(x-P\hat x),
\end{align}
where $P$ and $\tilde M\succ0$ are some matrices of appropriate dimensions. In order to show that $V$ in~\eqref{Eq_7a} is an SStF from $\widehat \Sigma_{\textsf{nl}}$ to $\Sigma_{\textsf{nl}}$, we require the following key assumption on $\Sigma_{\textsf{nl}}$.
\begin{assumption}\label{As_1a}
	Let $\Sigma_{\textsf{nl}}=(A,B,C_1,C_2,D,E,F,R,\varphi)$. Assume that for some constant
	$0<\widehat\kappa<1$ and $\tilde k > 0$ there exist matrices $\tilde M\succ0$, $K$, $L_1$, $Z$, $G$, $\bar X^{11}$, $\bar X^{12}$, $\bar X^{21}$, and $\bar X^{22}$ of appropriate dimensions such that the matrix equality
	\begin{align}\label{As_2a}
		D&= ZG,
	\end{align}
	and inequality~\eqref{Eq_8a} hold.
	
	\begin{figure*}
		\begin{align}\label{Eq_8a}
			&\begin{bmatrix}\notag
				(A+BK)^T\tilde M(A+BK) && (A+BK)^T\tilde MZ && (A+BK)^T\tilde M(BL_1+E) && (A+BK)^T\tilde M(B\tilde R-P\hat B)\\
				*&& Z^T \tilde MZ && Z^T \tilde M(BL_1+E) && Z^T \tilde M(B\tilde R-P\hat B) \\
				*&&*&&(BL_1+E)^T \tilde M(BL_1+E)&&(BL_1+E)^T \tilde M(B\tilde R-P\hat B) \\
				*&&  *&&*&&(B\tilde R-P\hat B)^T \tilde M(B\tilde R-P\hat B)
			\end{bmatrix}&\\
			&\preceq\begin{bmatrix}
				\widehat\kappa\tilde M+C_2^T\bar X^{22}C_2 & C_2^T\bar X^{21} & -F^T & 0\\
				\bar X^{12}C_2 & \bar X^{11} & 0 & 0\\
				-F & 0 & \frac{2}{b} & 0 \\ 0 & 0& 0& \tilde k (B\tilde R-P\hat B)^T \tilde M(B\tilde R-P\hat B)\\
			\end{bmatrix}
		\end{align}
		\rule{\textwidth}{0.1pt}
	\end{figure*}
\end{assumption}
Now, we provide one of the main results of this section showing under which conditions $V$ in~\eqref{Eq_7a} is an SStF from $\widehat \Sigma_{\textsf{nl}}$ to $\Sigma_{\textsf{nl}}$.
\begin{theorem}\label{Thm_3a}
	Let $\Sigma_{\textsf{nl}}=(A,B,C_1,C_2,D,E,F,R,\varphi)$
	and $\widehat \Sigma_{\textsf{nl}}=(\hat A,\hat B,\hat C_1,\hat C_2,\hat D,\hat E,\hat F,\hat R, \varphi)$ be two stochastic control subsystems with the same external output space dimension. Suppose Assumption~\ref{As_1a} holds and there exist matrices $P$, $Q$, $H$, $L_2$, and $\hat G$ such that
	\begin{IEEEeqnarray}{rCl}\IEEEyesnumber\label{Con_1056}
		\IEEEyessubnumber\label{Eq_10a} AP&=&P\hat A-BQ\\
		\IEEEyessubnumber\label{Eq_11a} C_1P&=&\hat C_1\\
		\IEEEyessubnumber\label{Eq_12a} \bar X^{12}C_2P&=&\bar X^{12}H\hat C_2\\
		\IEEEyessubnumber\label{Eq_13a} \bar X^{22}C_2P&=&\bar X^{22}H\hat C_2\\
		\IEEEyessubnumber\label{Eq_14a}  FP&=&\hat F\\
		\IEEEyessubnumber\label{Eq_15a}  E&=&P\hat E-B(L_1-L_2)\\
		\IEEEyessubnumber\label{Eq_16a}  P\hat D&=&Z\hat G,
	\end{IEEEeqnarray}	
	hold. Then, function $V$ defined in~\eqref{Eq_7a} is an SStF from $\widehat \Sigma_{\textsf{nl}}$ to $\Sigma_{\textsf{nl}}$.
\end{theorem}
The proof of Theorem~\ref{Thm_3a} is provided in the Appendix. 
Note that conditions~\eqref{Con_1056} hold as long as the geometric conditions V-$18$ to V-$23$ in~\cite{zamani2017compositionalMurat} hold. The functions $\alpha\in\mathcal{K}_\infty$, $\kappa\in\mathcal{K}$, $\rho_{\mathrm{ext}}\in\mathcal{K}_\infty\cup\{0\}$, and the matrix $\bar X$ in Definition~\ref{Def_1a} associated with the SStF in~\eqref{Eq_7a} are $\alpha(s)=\frac{\lambda_{\min}(\tilde M)}{\lambda_{\max}(C_1^TC_1)}s^2$\!, $\kappa(s):=(1-\widehat\kappa) s$, $\rho_{\mathrm{ext}}(s):=\tilde \kappa \Vert\sqrt{\tilde M}(B\tilde R-P\hat B)\Vert^2 s^2$\!, $\forall s\in\R_{\ge0}$, where $\tilde R$ is a matrix of appropriate dimension employed in the interface map~\eqref{Eq_405a}, and $\bar X=\begin{bmatrix}
\bar X^{11}&\bar X^{12}\\
\bar X^{21}&\bar X^{22}
\end{bmatrix}$. Moreover, positive constant $\psi$ in~\eqref{Eq_3a} is $\psi=\text{Tr}\big(R^T\tilde MR+\hat R^TP^T\tilde MP\hat R\big)$.

It is worth mentioning that for any linear system $\Sigma_{\textsf{nl}}=(A,B,C_1,C_2,D, R)$, stabilizability of the pair~$(A,B)$ is sufficient to satisfy Assumption~\ref{As_1a} in where matrices $E$, $F$, and $L_1$ are identically zero~\cite[Chapter 4]{antsaklis2007linear}.

One can readily verify from the result of Theorem~\ref{Thm_3a} that choosing $\hat R$ equal to zero results in smaller constant $\psi$ and, hence, more closeness between subsystems and their abstractions. This is not the case when one assumes the noise of the concrete subsystem and its abstraction are the same as in \cite{zamani2016approximations,zamani2014compositional}.

Since the results in Theorem~\ref{Thm_3a} do not impose any condition on matrix $\hat B$, it can be chosen arbitrarily. As an example, one can choose $\hat B=I_{\hat n}$ which makes the abstract system $\widehat \Sigma_{\textsf{nl}}$ fully actuated and, hence, the synthesis problem over it much easier.

\begin{remark}
	Since Theorem~\ref{Thm_3a} does not impose any condition on matrix $\tilde R$, one can choose $\tilde R$ such that it minimizes function $\rho_{\mathrm{ext}}$ for $V$ as suggested in \cite{girard2009hierarchical}. The following expression for $\tilde R$
	\begin{align}\notag
		\tilde R=(B^TMB)^{-1} B^TM P\hat B,
	\end{align}
	minimizes $\rho_{\mathrm{ext}}$.
\end{remark}

\section{Probability of Satisfaction for Properties Expressed as scLTL}
\label{Sec: Finite Automata}

Consider a dt-SCS $\Sigma\!=\!(\!X,U,\varsigma, \!f, \!Y,\!h)$ and a measurable target set $\mathsf B\subset Y$.
We say that an output trajectory $\{y(k)\}_{k\ge 0}$ reaches a target set $\mathsf B$ within time interval $[0,T_d]\subset \mathbb N$, if there exists a $k\in [0,T_d]$ such that $y(k)\in \mathsf B$.
This bounded reaching of $\mathsf B$ is denoted  by $ \lozenge^{\le T_d}\{y\in \mathsf B\}$ or briefly $\lozenge^{\le T_d}\mathsf B$.  For $T_d\rightarrow \infty$, we denote  the reachability property as $ \lozenge \mathsf B$, i.e., eventually $\mathsf B$.
For a dt-SCS $\Sigma$ with policy $\rho$,  we want to compute the probability that an output trajectory reaches $\mathsf B$ within the time horizon $T_d\in {\mathbb N}$, i.e., 
$\mathbb P(\lozenge^{\le T_d} \mathsf B)$.
The \emph{reachability probability} is the probability that the target set $\mathsf B$ is eventually reached and is denoted by 
$\mathbb P(\lozenge \mathsf B)$.

More complex properties can be described using temporal logic. Consider a set of atomic propositions $\AP$ and the alphabet $\alphabeth := 2^{\AP}$. 
Let $\word=\word(0),\word(1),\word(2),\ldots \in\alphabeth^{\mathbb{N}}$ be an infinite word, that is, a string composed of letters from $\alphabeth$.
Of interest are atomic propositions that are relevant to the dt-SCS via a measurable labeling function $\Lab$ from the output space to the alphabet
as
$\Lab:Y\rightarrow \alphabeth$. Output trajectories  $\{y(k)\}_{k\geq 0}\in Y^{\mathbb N} $ can be readily mapped to the set of infinite words $\alphabeth^{\mathbb N}$, as
\[\word=\Lab(\{y(k)\}_{k\geq0}):=\{\word\in \alphabeth^{\mathbb N}\,|\,\word(k)= \Lab(y(k)) \}.\]
Consider LTL properties with syntax \cite{baier2008principles}
\begin{equation*}
\phi ::=  \operatorname{true} \,|\, p \,|\, \neg \phi \,|\,\phi_1 \wedge \phi_2 \,|\, \nex \phi \,|\, \phi_1\until \phi_2.
\end{equation*}
Let $\word_k=\word(k),\word(k+1),\word(k+2),\ldots  $ be a subsequence (postfix) of $\word$, then  
the satisfaction relation between $\word$ and a property $\phi$, expressed in LTL,  is denoted by $\word\vDash\phi$  
(or equivalently $\word_0\vDash\phi$). 
The semantics of the satisfaction relation are defined recursively over $\word_k$ and
the syntax of the LTL formula $\phi$.
An atomic proposition $ p\in \AP$ is satisfied by $\word_k$, i.e.,  $\word_k\vDash p$, iff   $p \in\word(k)$.  Furthermore, 
$\word_k\vDash \neg \phi$  if $\word_k\nvDash\phi$ and 
we say that  $\word_k\vDash \phi_1\wedge\phi_2$ 
if $ \word_k\vDash \phi_1$ and $\word_k\vDash \phi_2$.
The next operator $\word_k\vDash\nex\phi $ holds if the property holds at the next time instance   $ \word_{k+1}\vDash \phi$. We denote by $\nex^j$, $j\in\N$, $j$ times composition of the next operator. With a slight abuse of the notation, one has $\nex^0\phi=\phi$ for any property $\phi$.
The temporal until operator $\word_k\vDash \phi_1\until\phi_2$  holds if $ \exists i \in \mathbb{N}:$ $\word_{k+i} \vDash \phi_2, \mbox{and } 
\forall j \in{\mathbb{N}:} 0\leq j<i, \word_{k+j}\vDash \phi_1
$.
Based on these semantics, operators such as disjunction ($\vee$) can also be defined through the negation and conjunction:
$ \word_k\vDash \phi_1\vee\phi_2\ \Leftrightarrow  \  \word_k\vDash \neg(\neg\phi_1 \wedge \neg\phi_2)$.

\begin{remark}
   Note that in this section, the satisfaction relation $\vDash$ changes by varying the labeling functions $\Lab$. In the following, we employ subscript for $\models$ to show its dependency on the labeling functions.
\end{remark}

We are interested in a fragment of LTL properties known as syntactically co-safe linear temporal logic (scLTL) \cite{KupfermanVardi2001}. This fragment is defined as follows.
\begin{definition}\label{def:scLTL}
	An scLTL over a set of atomic propositions $\AP$ has syntax 
	\begin{equation*}
	\phi ::=  \operatorname{true} \,|\, p \,|\, \neg p \,|\,\phi_1\wedge \phi_2\,|\,\phi_1 \vee \phi_2\,|\, \nex \phi \,|\, \phi_1\until \phi_2\,|\, \lozenge \phi,
	\end{equation*} 
	with $p\in \AP$.
\end{definition}

Even though scLTL formulas are defined over infinite words (as in LTL formulae), their satisfaction is guaranteed in finite time~\cite{KupfermanVardi2001}. Any infinite word $\word\in\alphabeth^{\mathbb{N}}$ satisfying an scLTL formula $\phi$ has a finite word $\word_f\in\alphabeth^n$, $n\in\mathbb N$, as its prefix such that all infinite words with prefix $\word_f$ also satisfy the formula $\phi$. We denote the language of such finite prefixes associated with an scLTL formula $\phi$ by $\mathcal L_f(\phi)$.

In the remainder, we consider scLTL properties since their verification can be performed via a reachability property over a finite state automaton \cite{KupfermanVardi2001,Calin17}. For this purpose, in this section we introduce a class of models known as Deterministic Finite-state Automata (DFA).
\begin{definition}
	A DFA is a tuple $\mathcal A = \left(Q_{\ell},q_0,\Sigma_{\textsf{a}},F_{\textsf{a}},\trans\right)$, where
	$Q_{\ell}$ is a finite set of locations,
	$q_0\in Q_{\ell}$ is the initial location,
	$\Sigma_{\textsf{a}}$ is a finite set (a.k.a. alphabet),
	$F_{\textsf{a}}\subseteq Q_{\ell}$ is a set of accept locations, and
	$\trans: Q_{\ell}\times\Sigma_{\textsf{a}}\rightarrow Q_{\ell}$ is a transition function.
\end{definition}

A finite word composed of letters of the alphabet, i.e., $\word_f = (\word_f(0),\ldots,\word_f(n))\in \Sigma_{\textsf{a}}^{n+1}$, is accepted by a DFA $\mathcal A$ if there exists a finite run $q =(q(0),\ldots,q(n + 1))\in Q_{\ell}^{n+2}$ such that $q(0) = q_0$,
$q(i + 1) =\trans(q(i),\word_f(i))$ for all $0\le i\le n$ and $q(n+1)\in F_{\textsf{a}}$. The accepted language of $\mathcal A$, denoted $\mathcal L(\mathcal A)$, is the set of all words accepted by $\mathcal A$. For every scLTL property $\phi$, cf. Definition \ref{def:scLTL}, there exists a DFA $\mathcal A_{\phi}$ such that
\begin{equation}
\mathcal L_f(\phi) = \mathcal L(\mathcal A_{\phi}).
\end{equation} 
As a result, the satisfaction of the property $\phi$ now becomes equivalent to the reaching of the accept locations in the DFA.
We use the DFA $\mathcal A$ to specify properties of dt-SCS $\Sigma=(X,U,\varsigma, f, Y,h)$ as follows.
Recall that $\mathsf L : Y\rightarrow \Sigma_{\textsf{a}}$ is a given measurable function. To each output $y\in Y$ it
assigns the letter $\mathsf L(y)\in\Sigma_{\textsf{a}}$. Given a policy $\rho$, we can define the probability that an output trajectory of $\Sigma$ satisfies an scLTL property $\phi$ over time horizon $[0,T_d]$, i.e. $\mathbb P(\word_f \in\mathcal L(\mathcal A_\phi)~\text{s.t.}~ |\word_f|\le T_d+1)$, with $|\word_f|$ denoting the length of $\word_f$ \cite{DLT08}.

\begin{figure}[ht]
	\begin{center}
		\includegraphics[width=7cm]{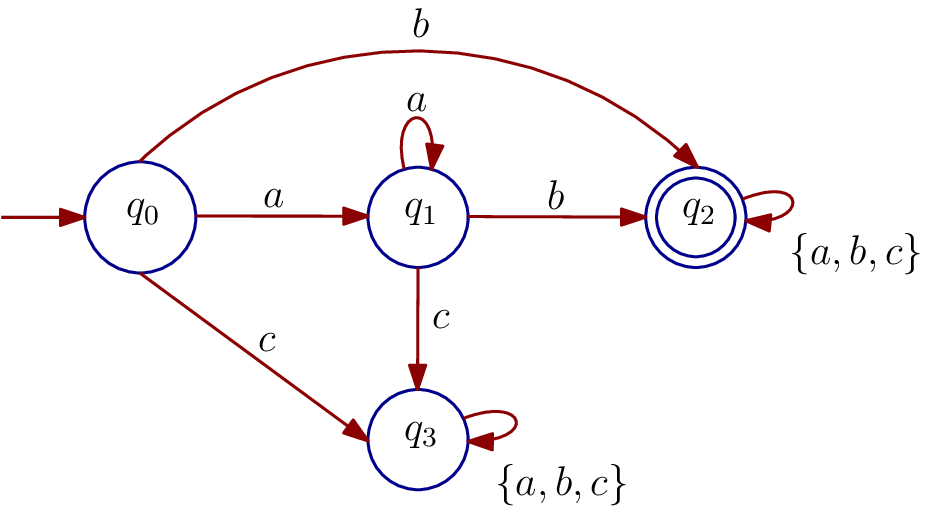}\hspace{1.3cm}
		\includegraphics[width=7cm]{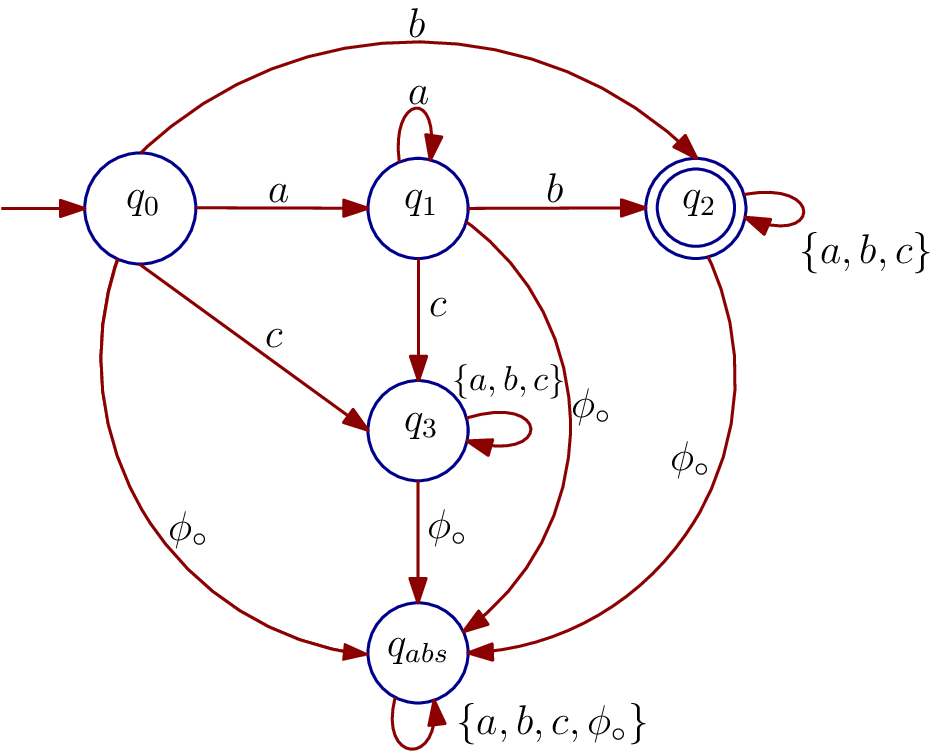}
		\caption{DFA $\mathcal A_{\phi}$ (left) and modified DFA $\mathcal A_{\hat\phi}$ (right) of the reach-avoid specification $(a\until b)$.}
		\label{Fig2}
	\end{center}
\end{figure}

The following example provides an automaton associated with a reach-avoid specification.
\begin{example}\label{illustration}
	Consider two measurable sets $\mathsf A,\mathsf B\subset Y$ as the safe and target sets, respectively. We present the DFA for the specification $(\mathsf A\until\mathsf B)$, which requires the output trajectories to reach the target set $\mathsf B$ while remaining in the safe set $\mathsf A$. Note that we do not assume these two sets being disjoint. Consider the set of atomic propositions $AP = \{\mathsf A,\mathsf B\}$ and the alphabet $\Sigma_{\textsf{a}} = \{\emptyset,\{\mathsf A\},\{\mathsf B\}, \{\mathsf A,\mathsf B\}\}$. Define the labeling function as
	\begin{equation*}
		\Lab(y) =
		\begin{cases}
			\{\mathsf A\}=:a & \text{ if }\,\, y\in \mathsf A\backslash \mathsf B,\\
			\{\mathsf B\}=:b & \text{ if }\,\,y\in \mathsf B,\\
			\emptyset=:c & \text{ if } \,\,y\notin\mathsf A\cup \mathsf B.
		\end{cases}
	\end{equation*}
	As can be seen from the above definition of the labeling function $\Lab$, it induces a partition over the output space $Y$ as
	\begin{equation*}
		\Lab^{-1}(a) = \mathsf A\backslash \mathsf B,\quad \Lab^{-1}(b) = \mathsf B,\quad \Lab^{-1}(c) = Y\backslash(\mathsf A\cup \mathsf B).
	\end{equation*}
	Note that we have indicated the elements of $\Sigma_{\textsf{a}}$ with lower-case letters for the ease of notation. 
	The specification $(\mathsf A\until\mathsf B)$ can be equivalently written as $(a\until b)$ with the associated DFA depicted in Figure~\ref{Fig2} left.
	This DFA has the set of locations $Q_{\ell}=\{q_0,q_1,q_2,q_3\}$, the initial location $q_0$, and accepting location $F_{\textsf{a}} =\{q_2\}$.
	Thus output trajectories of a dt-SCS $\Sigma$ satisfies the specification $(a\until b)$ if and only if their associated words are accepted by this DFA.
	
\end{example}

In the rest of this article, we focus on the computation of probability of $\omega \in\mathcal L(\mathcal A_\phi)$ over bounded intervals. In other words, we fix a time horizon $T_d$ and compute $\mathbb P(\omega(0)\omega(1)\ldots\omega(T_d)\in\mathcal L(\mathcal A_\phi))$.
Suppose $\Sigma$ and $\widehat\Sigma$ are two dt-SCS for which the results of Theorem~\ref{Thm_1a} hold.
Consider a labeling function $\Lab$ defined on their output space and an scLTL specification $\phi$ with DFA $\mathcal A_{\phi}$.
In the following we show how to construct a DFA $\mathcal A_{\hat\phi}$ of another specification $\hat\phi$ and a new labeling function $\Lab^\epsilon$ such that the satisfaction probability of $\hat\phi$ by output trajectories of $\widehat\Sigma$ and labeling function $\Lab^\epsilon$ gives a lower bounded on the satisfaction probability of $\phi$ by output trajectories of $\Sigma$ and labeling function $\Lab$.

Consider the labeling function $\Lab: Y\rightarrow \Sigma_{\textsf{a}}$. The new labeling function $\Lab^\epsilon : Y \rightarrow \bar\Sigma_{\textsf{a}}$ is constructed using the \mbox{$\epsilon$-perturbation} of subsets of $Y$. Define for any Borel measurable set $\mathsf A\subset Y$, its \mbox{$\epsilon$-perturbed} version $\mathsf A^\epsilon $ as the largest measurable set satisfying
\begin{equation*}
	\mathsf A^\epsilon \subseteq \{y\in\mathsf A\,\,\big|\,\,\|\bar y-y\|\geq\epsilon\, \text{ for all }\bar y\in Y\backslash\mathsf A\}.
\end{equation*}
Remark that the set $\mathsf A^\epsilon$ is just the largest measurable set contained in the $\epsilon$-deflated version of $A$ and without loss of generality we assume it is nonempty.
Then $\Lab^\epsilon(y) = \Lab(y)$ for any $y\in \cup_{a\in\Sigma_{\textsf{a}}}[\Lab^{-1}(a)]^\epsilon$, otherwise $\Lab^\epsilon(y) = \phi_\circ$.

Consider the DFA
$\mathcal A_{\phi} = \left(Q_{\ell},q_0,\Sigma_{\textsf{a}},F_{\textsf{a}},\trans\right)$.
The new DFA
\begin{equation}
	\label{eq:newDFA}
	\mathcal{A}_{\hat\phi}=(\bar Q_{\ell},q_0,\bar\Sigma_{\textsf{a}},F_{\textsf{a}},\bar{\trans})
\end{equation}
will be constructed by adding one absorbing location $q_{\textsf{abs}}$ and one letter $\phi_\circ$ as
$\bar Q_{\ell}:=Q_{\ell} \cup \{q_{\textsf{abs}}\}$
and $\bar \Sigma_{\textsf{a}}:=\Sigma_{\textsf{a}} \cup \{\phi_\circ\}$. The initial and accept locations are the same with $\mathcal A_{\phi}$.
The transition relation is defined, $\forall q \in \bar Q_{\ell}, \forall a \in \bar \Sigma_{\textsf{a}}$, as
\begin{equation*}
	\bar{\trans} (q, a):=
	\begin{cases}
		\trans (q,a)& \text{if}~ q \in Q_{\ell}, a \in \Sigma_{\textsf{a}},\\
		q_{\textsf{abs}} &\text{if}~ a = \phi_\circ, q \in \bar Q_{\ell}, \\
		q_{\textsf{abs}} & \text{if}~ q = q_{\textsf{abs}}, a \in \bar \Sigma_{\textsf{a}}.
	\end{cases}
\end{equation*}
In other words, we add an absorbing state $q_{\textsf{abs}}$ and all the states will jump to this absorbing state with label $\phi_\circ$.
As an example the modified DFA of the reach-avoid specification in Figure \ref{Fig2} left is plotted in Figure~\ref{Fig2} right.

In the next lemma, we employ the new labeling function to relate satisfaction of specifications by output trajectories of two dt-SCS.

\begin{lemma}
	\label{lem:ineq_DFA}
	Suppose two observed sequences of output trajectories for two dt-SCS $\Sigma$ and $\widehat\Sigma$ satisfy the inequality
	$$\sup_{0\le k\le T_d}\Vert y(k)-\hat y(k) \Vert < \epsilon,$$
	for some time bound $T_d$ and $\epsilon>0$.
	Then $y(\cdot)\vDash_\Lab\phi$ if $\hat y(\cdot)\vDash_{\Lab^\epsilon} \hat\phi$ over time interval $[0,T_d]$ with labeling functions $\Lab$ and $\Lab^\epsilon$ and modified specification $\hat\phi$ defined in \eqref{eq:newDFA}.
	
\end{lemma}
The proof of Lemma~\ref{lem:ineq_DFA} is provided in the Appendix. Next theorem presents the core result of this section.
\begin{theorem}\label{Thm: Lowerbound}
	Suppose $\Sigma$ and $\widehat\Sigma$ are two dt-SCS for which inequality \eqref{Eq_25} holds with the pair $(\epsilon,\delta)$ and any time bound $T_d$.
	Suppose a specification $\phi$ and a labeling function $\Lab$ is defined for $\Sigma$. The following inequality holds for the labeling function $\Lab^\epsilon$ on $\widehat\Sigma$ and modified specification $\hat\phi$:
	\begin{equation}
		\label{eq:lower_bound}
		\PP(\hat y(\cdot)\vDash_{\Lab^\epsilon} \hat\phi)-\delta \leq \PP (y(\cdot)\vDash_\Lab \phi),
	\end{equation}
	where the satisfaction is over time interval $[0,T_d]$.
\end{theorem}
The proof of Theorem~\ref{Thm: Lowerbound} is provided in the Appendix. 
In order to get an upper bound for $\PP (y(\cdot)\vDash_\Lab \phi)$, we need to define for any Borel measurable set $\mathsf A\subset Y$, its \mbox{$(-\epsilon)$-perturbed} version $\mathsf A^{-\epsilon}$ as the smallest measurable set satisfying 
\begin{equation*}
	\mathsf A^{-\epsilon} \supseteq \{y\in Y\,\big|\,\exists \bar y\in \mathsf A\text{ with } \|\bar y-y\|<\epsilon\}.
\end{equation*}
Remark that the set $\mathsf A^{-\epsilon}$ is just the smallest measurable set containing the \mbox{$\epsilon$-inflated} version of $\mathsf A$.

A new labeling map $\Lab^{-\epsilon}:Y\rightarrow 2^{\Sigma_{\textsf a}}$ is constructed using the \mbox{$(-\epsilon)-$perturbation} of subsets of $Y$ as
\begin{equation}
	\label{eq:lab_2}
	\Lab^{-\epsilon}(y) := \left\{a\in\Sigma_{\textsf{a}}\,|\, y\in [\Lab^{-1}(a)]^{-\epsilon}\right\}\!.
\end{equation}

\begin{theorem}
	Suppose $\Sigma$ and $\widehat\Sigma$ are two dt-SCS for which inequality \eqref{Eq_25} holds with the pair $(\epsilon,\delta)$ and any time bound $T_d$.
	Suppose a specification $\phi$ and a labeling function $\Lab$ is defined for $\Sigma$. The following inequality holds for the labeling function $\Lab^{-\epsilon}$ defined in \eqref{eq:lab_2} on $\widehat\Sigma$:
	\begin{equation}
		\label{eq:upper_bound}
		\PP (y(\cdot)\vDash_\Lab\phi)\le \PP(\hat y(\cdot)\vDash_{\Lab^{-\epsilon}}\phi)+\delta,
	\end{equation}
	where the satisfaction is over time interval $[0,T_d]$ and the probability in the right-hand side is computed for having $\hat y(\cdot)\vDash_{\Lab^{-\epsilon}}\phi$ for any choice of non-determinism introduced by the labeling map $\Lab^{-\epsilon}$.
\end{theorem}

The proof is similar to that of Theorem~\ref{Thm: Lowerbound}, and is omitted here due to lack of space.

In contrast with inequality \eqref{eq:lower_bound}, the specification $\phi$ is the same in both sides of \eqref{eq:upper_bound}. The non-determinism originating from $\mathsf{L}^{-\epsilon}$ in the right-hand side of \eqref{eq:upper_bound} can be pushed to the DFA representation of $\phi$, by constructing a finite automaton that is non-deterministic.

In the next section, we demonstrate the effectiveness of the proposed results for an interconnected system consisting of three nonlinear stochastic control subsystems in a compositional fashion.

\section{Case Study}\label{Sec: Case Study}
Consider a discrete-time nonlinear stochastic control system $\Sigma_{\textsf{nl}}$ satisfying
\begin{align}\notag
	\Sigma_{\textsf{nl}}:\left\{\hspace{-0.5mm}\begin{array}{l}x(k+1)=\bar Gx(k)+\varphi(x(k))+\nu(k)+R\varsigma(k),\\
		y(k)=Cx(k),\end{array}\right.
\end{align}
for some matrix $\bar G=(I_n-\tau L)\in\R^{n\times n}$ where $\tau L$ is the Laplacian matrix of an undirected graph with $0<\tau <1/\Delta$, where $\Delta$ is the maximum degree of the graph \cite{godsil2001}. Moreover, $R = \mathsf{diag}(0.007\mathbf{1}_{n_1},\ldots,0.007\mathbf{1}_{n_N})$, $\varsigma(k)=[\varsigma_1(k);\ldots;\varsigma_N(k)]$, $\varphi(x)=[\mathbf{1}_{n_1}\varphi_1(F_1 x_1(k));\ldots;\mathbf{1}_{n_N}\varphi_N(F_N x_N(k))]$ where $n=\sum_{i=1}^Nn_i$, $\varphi_i(x) = sin(x)$, and $F_i^T= \begin{bmatrix}1& 0& \cdots & 0\end{bmatrix}^T\in\R^{n_i}$ $\forall i\in\{1,\ldots,N\}$, and $C$ has the block diagonal structure as $C=\mathsf{diag}(C_{11},\ldots,C_{1N})$, where $C_{1i} \in\R^{q_i\times n_i}, \forall i\in\{1,\ldots,N\}$. We partition $x$ as $x=[x_1;\ldots;x_N]$ and $\nu$ as $\nu=[\nu_1;\ldots;\nu_N]$, where $x_i,\nu_i\in\R^{n_i}$. Now, by introducing $\Sigma_{\textsf{nl}i}=(I_{n_i}, I_{n_i}, C_{1i},I_{n_i},I_{n_i}, \mathbf{1}_{n_i},F_i, 0.007\mathbf{1}_{n_i},\varphi_i )$ satisfying
\begin{align}\notag
	\Sigma_{\textsf{nl}i}:\left\{\hspace{-1mm}\begin{array}{l}x_i(k+1)=x_i(k)+\mathbf{1}_{n_i}\varphi_i(F_i x_i(k))+\nu_i(k)+w_i(k)+0.007\mathbf{1}_{n_i}\varsigma_i(k),\\
		y_{1i}(k)=C_{1i}x_i(k),\\
		y_{2i}(k)=x_i(k),\end{array}\right.
\end{align}
one can readily verify that $\Sigma_{\textsf{nl}}=\mathcal{I}(\Sigma_{\textsf{nl}1},\ldots,\Sigma_{\textsf{nl}N})$ where the coupling matrix $M$ is given by $M=-\tau L$.
Our goal is to aggregate each $x_i$ into a scalar-valued $\hat x_i$, governed by $\widehat \Sigma_{\textsf{nl}i}=(0.5,1,\hat C_{1i}, 1,1, 0.1, 1, \varphi_i)$ which satisfies:
\begin{align}\notag
	\widehat \Sigma_{\textsf{nl}i}:\left\{\hspace{-1mm}\begin{array}{l}\hat x_i(k+1)=0.5\hat x_i(k)+0.1\varphi_i(\hat x_i(k))+\hat \nu_i(k)+\hat w_i(k),\\
		\hat y_{1i}(k)=\hat C_{1i}\hat x_i(k),\\
		\hat y_{2i}(k)=\hat x_i(k),\end{array}\right.
\end{align}
where $\hat C_{1i}=C_{1i}\mathbf{1}_{n_i}$. Note that here $\hat R_i$, $\forall i\in\{1,\ldots,N\}$, are considered zero in order to reduce constants $\psi_i$ for each $V_i$. One can readily verify that, for any $i\in\{1,\ldots,N\}$, conditions~\eqref{As_2a} and~\eqref{Eq_8a} are satisfied with $\tilde M_i=I_{n_i}$, $\widehat \kappa_i=0.95$, $\tilde \kappa_i=1$, $K_i=(\lambda_i-1) I_{n_i}$, $\lambda_i = 0.5$, $Z_i=G_i=I_{n_i}$, $L_{1i}=-\mathbf{1}_{n_i}$, $\tilde R =\mathbf{1}_{n_i}$, $\bar X^{11}=I_{n_i}$, $\bar X^{22}=0_{n_i}$, and $\bar X^{12}=\bar X^{21}=\lambda_i I_{n_i}$. Moreover, for any $i\in\{1,\ldots,N\}$, $P_i=\mathbf{1}_{n_i}$ satisfies conditions~\eqref{Con_1056} with $Q_i=-0.5\mathbf{1}_{n_i}$, ${L_2}_i=-0.1\mathbf{1}_{n_i}$, and $H_i=\hat G_i=\mathbf{1}_{n_i}$. Hence, function $V_i(x_i,\hat{x}_i)=(x_i-\mathbf{1}_{n_i}\hat x_i)^T(x_i-\mathbf{1}_{n_i}\hat x_i)$ is an SStF from $\widehat \Sigma_{\textsf{nl}i}$ to $\Sigma_{\textsf{nl}i}$ satisfying condition~\eqref{Eq_2a} with $\alpha_{i}(s)=\frac{1}{\lambda_{\max}(C_{1i}^TC_{1i})}s^2$ and condition~\eqref{Eq_3a} with $\kappa_i(s):=0.05s$, $\rho_{\mathrm{ext}i}(s)=0$, $\forall s\in\R_{\ge0}$, $G_i=I_{n_i}$, $H_i=\mathbf{1}_{n_i}$, and
\begin{equation}\label{Eq_22}
	\bar X_i=\begin{bmatrix} I_{n_i} & \lambda_i I_{n_i} \\ \lambda_i I_{n_i} &  0_{n_i} \end{bmatrix}\!\!,
\end{equation}
where the input $\nu_i$ is given via the interface function in~\eqref{Eq_405a} as
\begin{align}\notag
	\nu_i=(\lambda_i-1)(x_i-\mathbf{1}_{n_i}\hat x_i)- 0.5\mathbf{1}_{n_i}\hat x_i + \mathbf{1}_{n_i}\hat \nu_i- \mathbf{1}_{n_i} \varphi_i (F_i x_i)+ 0.1\mathbf{1}_{n_i} \varphi_i (F_i \mathbf{1}_{n_i} \hat x_i). 
\end{align} 
Now, we look at $\widehat \Sigma_{\textsf{nl}}=\mathcal{I}(\widehat \Sigma_{\textsf{nl}1},\ldots,\widehat \Sigma_{\textsf{nl}N})$ with a coupling matrix $\hat M$ satisfying condition~\eqref{Con_2a} as follows:
\begin{equation}\label{Eq_89}
	-\tau L~\mathsf{diag}(\!\mathbf{1}_{n_1},\ldots,\mathbf{1}_{n_N}\!)\!=\!\mathsf{diag}(\!\mathbf{1}_{n_1},\ldots,\mathbf{1}_{n_N}\!)\hat M.
\end{equation}
Note that the existence of $\hat M$ satisfying~\eqref{Eq_89} for graph Laplacian $\tau L$ means that the $N$ subgraphs form an {\it equitable partition} of the full graph \cite{godsil2001}. Although this restricts the choice of a partition in general, for the complete graph any partition is equitable.

Choosing $\mu_1=\cdots=\mu_N=1$ and using $\bar X_i$ in~\eqref{Eq_22}, matrix $\bar X_{cmp}$ in~\eqref{Def_3a} reduces to
$$
\bar X_{cmp}=\begin{bmatrix} I_{n} & \lambda I_{n} \\ \lambda I_{n} &  0_{n} \end{bmatrix}\!\!,
$$
where $\lambda=\lambda_1=\cdots=\lambda_N = 0.5$, and condition~\eqref{Con_1a} reduces to
\begin{align}\notag
	\begin{bmatrix}  -\tau L\\ I_n \end{bmatrix}^T\!\!\bar X_{cmp}\begin{bmatrix}  -\tau L \\ I_n \end{bmatrix}= \tau^2 L^T L-\lambda \tau L\notag-\lambda \tau L^T=\tau L(\tau L-2\lambda I_n)\preceq 0,
\end{align}
without requiring any restrictions on the number or gains of the subsystems with $\tau = 0.9/(n-1)$. In order to show the above inequality, we used $\tau L=\tau L^T\succeq0$ which is always true for Laplacian matrices of undirected graphs. Now, one can readily verify that $V(x,\hat x)=\sum_{i=1}^n(x_i-\mathbf{1}_{n_i}\hat x_i)^T(x_i-\mathbf{1}_{n_i}\hat x_i)$ is an SSF from  $\widehat\Sigma_{\textsf{nl}}$ to $\Sigma_{\textsf{nl}}$ satisfying conditions \eqref{eq:lowerbound2} and \eqref{eq:martingale2}.

For the sake of simulation, we assume $L$ is the Laplacian matrix of a complete graph. We fix $N=3$, $n=222$, $n_i=74$, and $C_{1i}=[1\,\, 0\,\, 0\,\, \ldots \,\, 0]$, $i\in \{1,2,3\}$. By using inequality~\eqref{Eq_25} and starting the interconnected systems  $\Sigma_{\textsf{nl}}$ and $\widehat\Sigma_{\textsf{nl}}$ from initial states $-13\mathbf{1}_{222}$ and $-13\mathbf{1}_{3}$, respectively, we guarantee that the distance between outputs of $\Sigma_{\textsf{nl}}$ and $\widehat \Sigma_{\textsf{nl}}$ will not exceed $\varepsilon = 1$ during the time horizon $T_d=10$ with probability at least $90\%$, i.e.
\begin{align}\notag
	\mathbb P\left(\Vert y_{a\nu}(k)-\hat y_{\hat a \hat\nu}(k)\Vert\le 1,\,\, \forall k\in[0,10]\right)\ge 0.9\,.
\end{align}

\begin{figure}
	\centering
	\includegraphics[width=10cm]{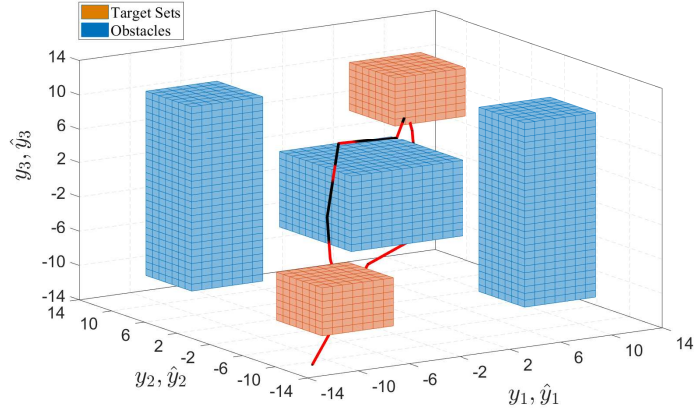}
	\caption{The specification with closed loop output trajectories of $\Sigma_{\textsf{nl}}$ (black one) and  $\widehat \Sigma_{\textsf{nl}}$ (red one). The sets $S$, $O_i$, $i\in\{1,2,3\}$, and $\bar T_i$, $i\in\{1,2\}$ are given by: $S=[-14,14]^3$,  $O_1=[-10,-6]\times[6,10]\times[10,10]$, $O_2=[-5,5]^3$, and $O_3=[6,10]\times[-10,-6]\times[10,10]$, $\bar T_1=[-10,-6]\times[-10,-6]\times[-10,-6]$ and $\bar T_2=[6,10]\times[6,10]\times[6,10]$.}
	\label{fig3}
\end{figure}

\begin{figure}
	\centering
	\includegraphics[width=9cm]{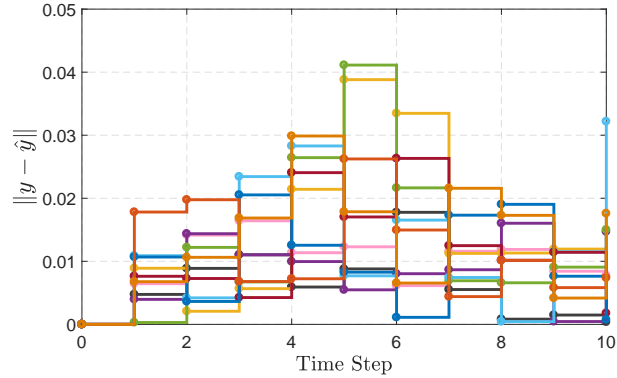}
	\caption{A few realizations of the norm of the error between the outputs of $\Sigma_{\textsf{nl}}$ and of $\widehat \Sigma_{\textsf{nl}}$, e.g. $\Vert y-\hat y \Vert$, for $T_d=10$.}
	\label{fig4}
\end{figure}

Let us now synthesize a controller for $\Sigma_{\textsf{nl}}$ via the abstraction $\widehat \Sigma_{\textsf{nl}}$ to enforce the specification, defined
by the following  scLTL formula (cf. Definition~\ref{def:scLTL}):
\begin{align}
	\varpi =& \underset{j=0}{\overset{T_d}{\bigwedge}} \nex^j\Big( S\wedge\Big( \underset{i=1}{\overset{3}{\bigwedge}} (\lnot O_i)\Big)\Big) \wedge \Diamond \bar T_1 \wedge  \Diamond \bar T_2,\label{LTL}
\end{align}
which requires that any output trajectory $y$ of the closed loop
system evolves inside the set $S$, avoids sets $O_i$, $i\in\{1,2,3\}$, indicated with blue boxes in Figure~\ref{fig3}, over bounded time interval $[0,T_d]$, and visits each $\bar T_i$, $i\in\{1,2\}$, indicated with red boxes in Figure~\ref{fig3}. We want to satisfy $\varpi$ over bounded time interval $[0,10]$ and take $T_d = 10$. We use \texttt{SCOTS} \cite{rungger2016scots} to
synthesize a controller for $\widehat \Sigma_{\textsf{nl}}$ to enforce~\eqref{LTL}. In the synthesis
process we restricted the abstract inputs $\hat \nu_1,\hat \nu_2,\hat \nu_3$ to $[-4,4]$. We also set the initial states of $\Sigma_{\textsf{nl}}$ to $x_i = P_i\hat x_i$, so that $V_i (x_i, \hat x_i) = 0$. A realization of closed-loop output trajectories of $\Sigma_{\textsf{nl}}$ and $\widehat  \Sigma_{\textsf{nl}}$ is illustrated in Figure~\ref{fig3}. Also, several realizations of the norm of error between outputs of
$\Sigma_{\textsf{nl}}$ and $\widehat \Sigma_{\textsf{nl}}$ are illustrated in Figure~\ref{fig4}. In order to have some more analysis on the provided probabilistic bound, we also run Monte Carlo simulation of $10000$ runs. In this case, one can statistically guarantee that the distance between outputs of $\Sigma_{\textsf{nl}}$ and $\widehat \Sigma_{\textsf{nl}}$  is always less than or equal to $0.04$ with the same probability, (i.e., at least $90\%$). This issue is expected and the reason is due to the conservatism nature of Lyapunov-like techniques (simulation function), but with the gain of having a formal guarantee on the output trajectories rather than empirical one. Note that it would not have been possible to synthesize a controller using \texttt{SCOTS} for the original $222$-dimensional system $\Sigma_{\textsf{nl}}$, without the $3$-dimensional intermediate approximation $\widehat \Sigma_{\textsf{nl}}$. We have intentionally dropped the noise of the abstraction and employed SCOTS here to show that if the concrete system possess some nice stability property and the noises of two systems are additive and independent, it is actually better to construct and use the non-stochastic abstraction since the non-stochastic abstraction is closer that the stochastic version (as discussed in Section~\ref{Sec: Nonlinear Control}).

\section{Discussion}
In this paper, we provided a compositional approach for infinite abstractions of interconnected discrete-time stochastic control systems, with independent noises in the abstract and concrete domains. To do so, we leveraged the interconnection matrix and joint dissipativity-type properties of subsystems and their abstractions. We introduced new notions of stochastic storage and simulation functions in order to quantify the distance in probability between original stochastic control subsystems and their abstractions and their interconnections, respectively. Using those notions, one can employ the proposed results here to synthesize policies enforcing certain temporal logic properties over abstract systems and then refine them to the ones for the concrete systems while quantifying the satisfaction errors in this detour process. We also provided a computational scheme for a class of discrete-time nonlinear stochastic control systems to construct their abstractions together with their corresponding stochastic storage functions. 
Furthermore, we addressed a fragment of LTL known as syntactically co-safe LTL, and showed how to quantify the probability of satisfaction for such specifications. Finally, we demonstrated the effectiveness of the results by constructing an abstraction (totally 3 dimensions) of the interconnection of three discrete-time nonlinear stochastic control subsystems (together 222 dimensions) in a compositional fashion. We also employed the abstraction as a substitute to synthesize a controller enforcing a syntactically co-safe LTL specification.

\bibliographystyle{alpha}
\bibliography{biblio}

\newcommand{\etalchar}[1]{$^{#1}$}
\begin{thebibliography}{ZMEM{\etalchar{+}}14}

\bibitem[AM07]{antsaklis2007linear}
P.~J. Antsaklis and A.~N. Michel.
\newblock {\em A linear systems primer}, volume~1.
\newblock Birkh{\"a}user Boston, 2007.

\bibitem[AMP16]{2016Murat}
M.~Arcak, C.~Meissen, and A.~Packard.
\newblock {\em Networks of dissipative systems}.
\newblock SpringerBriefs in Electrical and Computer Engineering. Springer,
  2016.

\bibitem[APLS08]{APLS08}
A.~Abate, M.~Prandini, J.~Lygeros, and S.~Sastry.
\newblock Probabilistic reachability and safety for controlled discrete time
  stochastic hybrid systems.
\newblock {\em Automatica}, 44(11):2724--2734, 2008.

\bibitem[BKL08]{baier2008principles}
C.~Baier, J.~P.r Katoen, and K.~G. Larsen.
\newblock {\em Principles of model checking}.
\newblock MIT press, 2008.

\bibitem[BS96]{BS96}
D.~P. Bertsekas and S.~E. Shreve.
\newblock {\em Stochastic {O}ptimal {C}ontrol: {T}he {D}iscrete-{T}ime {C}ase}.
\newblock Athena Scientific, 1996.

\bibitem[BYG17]{Calin17}
Calin Belta, Boyan Yordanov, and Ebru~Aydin Gol.
\newblock {\em Formal Methods for Discrete-Time Dynamical Systems}, volume~89
  of {\em Studies in Systems, Decision and Control}.
\newblock Springer International Publishing, 2017.

\bibitem[DLT08]{DLT08}
J.~Desharnais, F.~Laviolette, and M.~Tracol.
\newblock Approximate analysis of probabilistic processes: logic, simulation
  and games.
\newblock In {\em Proceedings of the International Conference on Quantitative
  Evaluation of SysTems (QEST 08)}, pages 264--273, September 2008.

\bibitem[GP09]{girard2009hierarchical}
A.~Girard and G.~J. Pappas.
\newblock Hierarchical control system design using approximate simulation.
\newblock {\em Automatica}, 45(2):566--571, 2009.

\bibitem[GR01]{godsil2001}
C.~Godsil and G.~Royle.
\newblock {\em Algebraic graph theory}.
\newblock Graduate Texts in Mathematics. Springe, 2001.

\bibitem[HS18]{HS19}
Sofie Haesaert and Sadegh Soudjani.
\newblock Robust dynamic programming for temporal logic control of stochastic
  systems.
\newblock {\em CoRR}, abs/1811.11445, 2018.

\bibitem[HSA17]{SSA16}
Sofie Haesaert, Sadegh {Soudjani}, and Alessandro Abate.
\newblock Verification of general {M}arkov decision processes by approximate
  similarity relations and policy refinement.
\newblock In {\em SIAM Journal on Control and Optimization}, volume~55, pages
  2333--2367, 2017.

\bibitem[JP09]{julius2009approximations}
A.~A. Julius and G.~J. Pappas.
\newblock Approximations of stochastic hybrid systems.
\newblock {\em IEEE Transactions on Automatic Control}, 54(6):1193--1203, 2009.

\bibitem[Kus67]{1967stochastic}
H.~J. Kushner.
\newblock {\em Stochastic Stability and Control}.
\newblock Mathematics in Science and Engineering. Elsevier Science, 1967.

\bibitem[KV01]{KupfermanVardi2001}
O~Kupferman and M~Y Vardi.
\newblock Model checking of safety properties.
\newblock {\em Formal Methods in System Design}, 19(3):291--314, 2001.

\bibitem[LSMZ17]{lavaei2017compositional}
A.~Lavaei, S.~{Soudjani}, R.~Majumdar, and M.~Zamani.
\newblock Compositional abstractions of interconnected discrete-time stochastic
  control systems.
\newblock In {\em Proceedings of the 56th IEEE Conference on Decision and
  Control}, pages 3551--3556, 2017.

\bibitem[LSZ18a]{lavaei2018ADHSJ}
A.~{Lavaei}, S.~{Soudjani}, and M.~{Zamani}.
\newblock Compositional (in)finite abstractions for large-scale interconnected
  stochastic systems.
\newblock {\em arXiv: 1808.00893}, August 2018.

\bibitem[LSZ18b]{lavaei2018ADHS}
A.~Lavaei, S.~{Soudjani}, and M.~Zamani.
\newblock Compositional synthesis of finite abstractions for continuous-space
  stochastic control systems: A small-gain approach.
\newblock In {\em Proceedings of the 6th IFAC Conference on Analysis and Design
  of Hybrid Systems}, volume~51, pages 265--270, 2018.

\bibitem[LSZ18c]{lavaei2017HSCC}
A.~Lavaei, S.~Soudjani, and M.~Zamani.
\newblock From dissipativity theory to compositional construction of finite
  {M}arkov decision processes.
\newblock In {\em Proceedings of the 21st ACM International Conference on
  Hybrid Systems: Computation and Control}, pages 21--30, 2018.

\bibitem[LSZ19]{lavaei2019NAHS}
A.~Lavaei, S.~Soudjani, and M.~Zamani.
\newblock Compositional synthesis of large-scale stochastic systems: A relaxed
  dissipativity approach.
\newblock {\em arXiv:1902.01223v2}, February 2019.

\bibitem[MSSM17]{2017arXiv170909546M}
K.~{Mallik}, S.~{Soudjani}, A.-K. {Schmuck}, and R.~{Majumdar}.
\newblock {Compositional Construction of Finite State Abstractions for
  Stochastic Control Systems}.
\newblock {\em arXiv: 1709.09546}, September 2017.

\bibitem[RZ16]{rungger2016scots}
M.~Rungger and M.~Zamani.
\newblock {SCOTS}: A tool for the synthesis of symbolic controllers.
\newblock In {\em Proceedings of the 19th ACM International Conference on
  Hybrid Systems: Computation and Control}, pages 99--104, 2016.

\bibitem[SA13]{SA13}
S.~{Soudjani} and A.~Abate.
\newblock Adaptive and sequential gridding procedures for the abstraction and
  verification of stochastic processes.
\newblock {\em SIAM Journal on Applied Dynamical Systems}, 12(2):921--956,
  2013.

\bibitem[SAM15]{SAM15}
S.~{Soudjani}, A.~Abate, and R.~Majumdar.
\newblock Dynamic {B}ayesian networks as formal abstractions of structured
  stochastic processes.
\newblock In {\em Proceedings of the 26th International Conference on
  Concurrency Theory}, pages 1--14, 2015.

\bibitem[SGA15]{FAUST15}
S.~{Soudjani}, C.~Gevaerts, and A.~Abate.
\newblock \textsf{FAUST}$^{\textsf{2}}$: Formal abstractions of
  uncountable-state stochastic processes.
\newblock In {\em TACAS'15}, volume 9035 of {\em Lecture Notes in Computer
  Science}, pages 272--286. Springer, 2015.

\bibitem[{Sou}14]{SSoudjani}
S.~{Soudjani}.
\newblock {\em Formal Abstractions for Automated Verification and Synthesis of
  Stochastic Systems}.
\newblock PhD thesis, Technische Universiteit Delft, The Netherlands, 2014.

\bibitem[TA11]{tkachev2011infinite}
I.~Tkachev and A.~Abate.
\newblock On infinite-horizon probabilistic properties and stochastic
  bisimulation functions.
\newblock In {\em Proceedings of the 50th IEEE Conference on Decision and
  Control and European Control Conference (CDC-ECC)}, pages 526--531, 2011.

\bibitem[Tab09]{tabuada2009verification}
P.~Tabuada.
\newblock {\em Verification and control of hybrid systems: a symbolic
  approach}.
\newblock Springer Science \& Business Media, 2009.

\bibitem[TMKA13]{tmka2013}
I.~Tkachev, A.~Mereacre, {J.-P.} Katoen, and A.~Abate.
\newblock Quantitative automata-based controller synthesis for non-autonomous
  stochastic hybrid systems.
\newblock In {\em Proceedings of the 16th ACM International Conference on
  Hybrid Systems: Computation and Control}, pages 293--302, 2013.

\bibitem[ZA14]{zamani2014approximately}
M.~Zamani and A.~Abate.
\newblock Approximately bisimilar symbolic models for randomly switched
  stochastic systems.
\newblock {\em Systems \& Control Letters}, 69:38--46, 2014.

\bibitem[ZA17]{zamani2017compositionalMurat}
M.~Zamani and M.~Arcak.
\newblock Compositional abstraction for networks of control systems: A
  dissipativity approach.
\newblock {\em IEEE Transactions on Control of Network Systems}, 2017.

\bibitem[ZAG15]{zamani2015symbolic}
M.~Zamani, A.~Abate, and A.~Girard.
\newblock Symbolic models for stochastic switched systems: A discretization and
  a discretization-free approach.
\newblock {\em Automatica}, 55:183--196, 2015.

\bibitem[Zam14]{zamani2014compositional}
M.~Zamani.
\newblock Compositional approximations of interconnected stochastic hybrid
  systems.
\newblock In {\em Proceedings of the 53rd IEEE Conference on Decision and
  Control (CDC)}, pages 3395--3400, 2014.

\bibitem[ZMEM{\etalchar{+}}14]{zamani2014symbolic}
M.~Zamani, P.~Mohajerin~Esfahani, R.~Majumdar, A.~Abate, and J.~Lygeros.
\newblock Symbolic control of stochastic systems via approximately bisimilar
  finite abstractions.
\newblock {\em IEEE Transactions on Automatic Control}, 59(12):3135--3150,
  2014.

\bibitem[ZRME17]{zamani2016approximations}
M.~Zamani, M.~Rungger, and P.~Mohajerin~Esfahani.
\newblock Approximations of stochastic hybrid systems: A compositional
  approach.
\newblock {\em IEEE Transactions on Automatic Control}, 62(6):2838--2853, 2017.

\end{thebibliography}

\section{Appendix}
\begin{proof}\textbf{(Theorem~\ref{Thm_1a})}
	Since $V$ is an SSF from $\widehat\Sigma$ to $\Sigma$, we have
	\begin{align}
		\PP&\left\{\sup_{0\leq k\leq T_d}\Vert y_{a\nu}(k)-\hat y_{\hat a \hat\nu}(k)\Vert\geq\varepsilon\,|\,[a;\hat a]\right\}
		=\PP\left\{\sup_{0\leq k\leq T_d}\alpha\left(\Vert y_{a\nu}(k)-\hat y_{\hat a \hat\nu}(k)\Vert\right)\geq\alpha(\varepsilon)\,|\,[a;\hat a]\right\}\nonumber\\
		&\leq\PP\left\{\sup_{0\leq k\leq T_d}V\left(x_{a\nu}(k),\hat x_{\hat a \hat\nu}(k)\right)\geq\alpha(\varepsilon)\,|\,[a;\hat a]\right\}.\label{eq:supermart}
	\end{align}
	The equality holds due to $\alpha$ being a $\mathcal K_\infty$ function. The inequality  is also true due to condition~\eqref{eq:lowerbound2} on the SSF $V$. The results follow by applying the first part of Lemma~\ref{Lemma: Kushner} to~\eqref{eq:supermart} with some slight modification and utilizing inequality~\eqref{eq:martingale2}.
	
\end{proof}	
\begin{proof}\textbf{(Corollary~\ref{corollary: Supermartingle})}
	Since $V$ is an SSF from $\widehat\Sigma$ to $\Sigma$ with $\rho_{\mathrm{ext}}(\cdot)\equiv0$ and $\psi = 0$, for any $x(k)\in X$ and $\hat x(k)\in\hat X$ and any $\hat\nu(k)\in\hat U$, there exists $\nu(k)\in U$ such that
	\begin{align}\notag
		\EE \Big[V(x(k+1),\hat x(k+1)\,|\,x(k),\hat x(k),\nu(k), \hat \nu(k)\Big]-V((x(k),\hat x(k))\leq-\kappa (V(x(k),\hat x(k)),
	\end{align}
	implying that $V\left(x_{a\nu}(k),\hat x_{\hat a \hat\nu}(k)\right)$ is a nonnegative supermartingale~\cite[Chapter 1]{1967stochastic} for any initial condition $a$ and $\hat a$ and inputs $\nu,\hat\nu$.
	Following the same reasoning as in the proof of Theorem~\ref{Thm_1a} we have
	\begin{align}\nonumber
		\PP&\left\{\sup_{0\leq k< \infty}\Vert y_{a\nu}(k)-\hat y_{\hat a \hat \nu}(k)\Vert\ge \varepsilon\,|\,[a;\hat a]\right\}
		=\PP\left\{\sup_{0\leq k< \infty}\alpha\Big(\Vert y_{a\nu}(k)-\hat y_{\hat a \hat \nu}(k)\Vert\Big)\ge \alpha(\varepsilon)\,|\,[a;\hat a]\right\}\\\notag
		&\leq\PP\left\{\sup_{0\leq k< \infty}V(x_{a\nu}(k),\hat x_{\hat a\hat \nu}(k))\ge \alpha(\varepsilon)\,|\,[a;\hat a]\right\}\leq\frac{V(a,\hat a)}{\alpha(\varepsilon)},
	\end{align}
	where the last inequality is due to the nonnegative supermartingale property as presented in the second part of Lemma~\ref{Lemma: Kushner}.
\end{proof}

\begin{proof}\textbf{(Theorem~\ref{Thm_2a})}
	We first show that inequality~\eqref{eq:lowerbound2} holds for some $\mathcal{K}_\infty$ function $\alpha$. For any $x=[{x_1;\ldots;x_N}]\in X$ and  $\hat x=[{\hat x_1;\ldots;\hat x_N}]\in \hat X$, one gets:
	\begin{align}\notag
		\Vert h(x)-\hat h(\hat x) \Vert&=\Vert [h_{11}(x_1);\ldots;h_{1N}(x_N)]-[\hat h_{11}(\hat x_1);\ldots;\hat h_{1N}(\hat x_N)]\Vert\\\notag
		&\le\sum_{i=1}^N \Vert  h_{1i}(x_i)-\hat h_{1i}(\hat x_i) \Vert
		\le \sum_{i=1}^N \alpha_{i}^{-1}(V_i( x_i, \hat x_i))\le \bar\alpha(V(x,\hat x)),
	\end{align} 
	with function $\bar\alpha:\mathbb R_{\ge 0}\rightarrow\mathbb R_{\ge 0}$ defined for all $r\in\mathbb R_{\ge 0}$ as
	\begin{center}
		$\bar\alpha(r) \Let \max\left\{\sum_{i=1}^N\alpha_{i}^{-1}(s_i)\,\,\big|\, s_i  {\ge 0},\,\,\sum_{i=1}^N \mu_i s_i=r\right\}. $
	\end{center}
	It is not hard to verify that function $\bar\alpha(\cdot)$ defined above is a $\mathcal{K}_\infty$ function.
	By taking the $\mathcal{K}_\infty$ function $\alpha(r):=\bar\alpha^{-1}(r)$, $\forall r\in\R_{\ge0}$, one obtains
	$$\alpha(\Vert h(x)-\hat h(\hat x)\Vert)\le V( x, \hat x),$$satisfying inequality~\eqref{eq:lowerbound2}.
	Now we prove that function $V$ in~\eqref{eq565} satisfies inequality~\eqref{eq:martingale2}. Consider any $x=[{x_1;\ldots;x_N}]\in X$, $\hat x=[{\hat x_1;\ldots;\hat x_N}]\in \hat X$, and
	$\hat \nu=[{\hat \nu_{1};\ldots;\hat \nu_{N}}]\in\hat U$. For any $i\in\{1,\ldots,N\}$, there exists $\nu_i\in U_i$, consequently, a vector $\nu=[{\nu_{1};\ldots;\nu_{N}}]\in U$, satisfying~\eqref{Eq_3a} for each pair of subsystems $\Sigma_i$ and $\widehat \Sigma_i$
	with the internal inputs given by $[{w_1;\ldots;w_N}]=M[h_{21}(x_1);\ldots;h_{2N}(x_N)]$ and $[{\hat w_1;\ldots;\hat w_N}]=\hat M[\hat h_{21}(\hat x_1);\ldots;\hat h_{2N}(\hat x_N)]$.
	Then we have the chain of inequalities in~\eqref{Eq_4a} using conditions~\eqref{Con_1a} and~\eqref{Con_2a} and by defining $\kappa(\cdot),\rho_{\mathrm{ext}}(\cdot),\psi$ as
	\begin{align}\notag
		\kappa(r) &\Let \min\left\{\sum_{i=1}^N\mu_i\kappa_i(s_i)\,\,\big|\, s_i  {\ge 0},\,\,\sum_{i=1}^N \mu_i s_i=r\right\} \\\notag
		\rho_{\mathrm{ext}}(r) &\Let \max\left\{\sum_{i=1}^N\mu_i\rho_{\mathrm{ext}i}(s_i)\big| s_i  {\ge 0},\|[{s_1;\ldots;s_N}]\| = r\right\}\\\notag
		\psi&\Let\sum_{i=1}^N\mu_i\psi_i.
	\end{align}
	Note that $\kappa$ and $\rho_{\mathrm{ext}}$ in~\eqref{Eq_4a} belong to $\mathcal{K}$ and $\mathcal{K}_\infty\cup\{0\}$, respectively, because of their definition provided above. Hence, we conclude that $V$ is an SSF from $\widehat \Sigma$ to $\Sigma$.
\end{proof}
\begin{figure*}[ht]
	\rule{\textwidth}{0.1pt}
	\begin{align}\notag\label{Eq_4a}
	\EE&\Big[\sum_{i=1}^N\mu_iV_i(x_i(k+1),\hat x_i(k+1))\,|\,[x(k)=x,\hat x(k)= \hat x,\hat{\nu}(k) = \hat \nu]\Big]-\sum_{i=1}^N\mu_iV_i(x_i,\hat x_i)\\\notag
	&= \sum_{i=1}^N\mu_i\EE\Big[V_i(x_i(k+1),\hat x_i(k+1))\,|\,[x_i(k) = x_i,\hat x_i(k) = \hat x_i,\hat \nu_i(k) = \hat \nu_i]\Big]-\sum_{i=1}^N\mu_iV_i(x_i,\hat x_i)\\\notag
	&\leq\sum_{i=1}^N\mu_i\bigg(\!-\kappa_i(V_i( x_i,\hat x_i))+\rho_{\mathrm{ext}i}(\Vert \hat \nu_i\Vert)+\psi_i+\begin{bmatrix}
	G_iw_i-\hat G_i\hat w_i\\
	h_{2i}(x_i)-H_i\hat h_{2i}(\hat x_i)
	\end{bmatrix}^T\begin{bmatrix}
	\bar X_i^{11}&\bar X_i^{12}\\
	\bar X_i^{21}&\bar X_i^{22}
	\end{bmatrix}\begin{bmatrix}
	G_iw_i-\hat G_i\hat w_i\\
	h_{2i}(x_i)-H_i\hat h_{2i}(\hat x_i)
	\end{bmatrix}\!\bigg)\\\notag
	&=\sum_{i=1}^N-\mu_i\kappa_i(V_i( x_i,\hat x_i))+\sum_{i=1}^N\mu_i\rho_{\mathrm{ext}i}(\Vert \hat \nu_i\Vert)+\sum_{i=1}^N\mu_i\psi_i\\\notag
	&+\begin{bmatrix}
	G_1w_1-\hat G_1\hat w_1\\
	\vdots\\
	G_Nw_N-\hat G_N\hat w_N\\
	h_{21}(x_1)-H_1\hat h_{21}(\hat x_1)\\
	\vdots\\
	h_{2N}(x_N)-H_N\hat h_{2N}(\hat x_N)
	\end{bmatrix}^T\begin{bmatrix}
	\mu_1\bar X_1^{11}&&&\mu_1\bar X_1^{12}&&\\
	&\ddots&&&\ddots&\\
	&&\mu_N\bar X_N^{11}&&&\mu_N\bar X_N^{12}\\
	\mu_1\bar X_1^{21}&&&\mu_1\bar X_1^{22}&&\\
	&\ddots&&&\ddots&\\
	&&\mu_N\bar X_N^{21}&&&\mu_N\bar X_N^{22}
	\end{bmatrix}\begin{bmatrix}
	G_1w_1-\hat G_1\hat w_1\\
	\vdots\\
	G_Nw_N-\hat G_N\hat w_N\\
	h_{21}(x_1)-H_1\hat h_{21}(\hat x_1)\\
	\vdots\\
	h_{2N}(x_N)-H_N\hat h_{2N}(\hat x_N)
	\end{bmatrix}\\\notag
	&=\sum_{i=1}^N-\mu_i\kappa_i(V_i( x_i,\hat x_i))+\sum_{i=1}^N\mu_i\rho_{\mathrm{ext}i}(\Vert \hat \nu_i\Vert)+\sum_{i=1}^N\mu_i\psi_i\\\notag
	&+\begin{bmatrix}
	GM\begin{bmatrix}
	h_{21}(x_1)\\
	\vdots\\
	h_{2N}(x_N)
	\end{bmatrix}\!-\!\hat G \hat M\begin{bmatrix}
	\hat h_{21}(\hat x_1)\\
	\vdots\\
	\hat h_{2N}(\hat x_N)
	\end{bmatrix}\\
	h_{21}(x_1)\!-\!H_1\hat h_{21}(\hat x_1)\\
	\vdots\\
	h_{2N}(x_N)\!-\!H_N\hat h_{2N}(\hat x_N)
	\end{bmatrix}^T\!\!\!\!\bar X_{cmp}\!\begin{bmatrix}
	GM\begin{bmatrix}
	h_{21}(x_1)\\
	\vdots\\
	h_{2N}(x_N)
	\end{bmatrix}\!-\!\hat G \hat M\begin{bmatrix}
	\hat h_{21}(\hat x_1)\\
	\vdots\\
	\hat h_{2N}(\hat x_N)
	\end{bmatrix}\\
	h_{21}(x_1)\!-\!H_1\hat h_{21}(\hat x_1)\\
	\vdots\\
	h_{2N}(x_N)\!-\!H_N\hat h_{2N}(\hat x_N)
	\end{bmatrix}\!=\!\sum_{i=1}^N\!-\mu_i\kappa_i(V_i( x_i,\hat x_i))\\\notag
	&+\!\sum_{i=1}^N\mu_i\rho_{\mathrm{ext}i}(\Vert \hat \nu_i\Vert)\!+\!\sum_{i=1}^N\mu_i\psi_i\!+\!\begin{bmatrix}
	h_{21}(x_1)-H_1\hat h_{21}(\hat x_1)\\
	\vdots\\
	h_{2N}(x_N)-H_N\hat h_{2N}(\hat x_N)
	\end{bmatrix}^T\!\begin{bmatrix}
	GM\\
	I_{\tilde q}
	\end{bmatrix}^T\!\bar X_{cmp}\begin{bmatrix}
	GM\\
	I_{\tilde q}
	\end{bmatrix}\!\begin{bmatrix}
	h_{21}(x_1)-H_1\hat h_{21}(\hat x_1)\\
	\vdots\\
	h_{2N}(x_N)-H_N\hat h_{2N}(\hat x_N)
	\end{bmatrix}\\
	&\leq\sum_{i=1}^N-\mu_i\kappa_i(V_i( x_i,\hat x_i))+\sum_{i=1}^N\mu_i\rho_{\mathrm{ext}i}(\Vert \hat \nu_i\Vert)+\sum_{i=1}^N\mu_i\psi_i\leq-\kappa\left(V\left( x,\hat{x}\right)\right)+\rho_{\mathrm{ext}}(\left\Vert \hat \nu\right\Vert)+\psi.
	\end{align}
	\rule{\textwidth}{0.1pt}
\end{figure*}
\begin{proof}\textbf{(Theorem~\ref{Thm_3a})}
	Here, we first show that $\forall x$, $\forall \hat x$, $\forall \hat \nu$, $\exists\nu$,  $\forall \hat w$, and $\forall w$, $V$ satisfies $\frac{\lambda_{\min}(\tilde M)}{\lambda_{\max}(C_1^TC_1)}\Vert C_1x-\hat C_1\hat x\Vert^2\le V(x,\hat x)$ and then
	\begin{align}\notag
		\EE& \Big[V(x(k+1),\hat x(k+1)\,|\,x(k) = x,\hat x(k) = \hat x, w(k)\!=\!w,\hat w(k)\!=\!\hat w,\hat \nu(k)\!=\! \hat \nu \Big]-V(x,\hat x)\\\notag 
		&\leq-(1-\widehat\kappa) (V(x,\hat x))+\tilde k\Vert\sqrt{\tilde M}(B\tilde R-P\hat B)\Vert^2\Vert\hat \nu\Vert^2+\begin{bmatrix}
			Gw-\hat G\hat w\\
			h_2(x)-H\hat h_2(\hat x)
		\end{bmatrix}^T{\begin{bmatrix}
				\bar X^{11}&\bar X^{12}\\
				\bar X^{21}&\bar X^{22}
		\end{bmatrix}}\begin{bmatrix}
			Gw-\hat G\hat w\\
			h_2(x)-H\hat h_2(\hat x)
		\end{bmatrix}\\\notag
		&+\text{Tr}\Big(R^T\tilde MR+\hat R^TP^T\tilde MP\hat R\Big).\notag
	\end{align}
	According to~\eqref{Eq_11a}, we have $\Vert C_1x-\hat C_1\hat x\Vert^2=(x-P\hat x)^TC_1^TC_1(x-P\hat x)$. Since  $\lambda_{\min}(C_1^TC_1)\Vert x- P\hat x\Vert^2\leq(x-P\hat x)^TC_1^TC_1(x-P\hat x)\leq\lambda_{\max}(C_1^TC_1)\Vert x- P\hat x\Vert^2$ and similarly  $\lambda_{\min}(\tilde M)\Vert x- P\hat x\Vert^2\leq(x-P\hat x)^T\tilde M(x-P\hat x)\leq\lambda_{\max}(\tilde M)\Vert x- P\hat x\Vert^2$, it can be readily verified that  $\frac{\lambda_{\min}(\tilde M)}{\lambda_{\max}(C_1^TC_1)}\Vert C_1x-\hat C_1\hat x\Vert^2\le V(x,\hat x)$ holds $\forall x$, $\forall \hat x$, implying that inequality~\eqref{Eq_2a} holds with $\alpha(s)=\frac{\lambda_{\min}(\tilde M)}{\lambda_{\max}(C_1^TC_1)}s^2$ for any $s\in\R_{\geq0}$. We proceed with showing that the inequality~\eqref{Eq_3a} holds, as well. Given any $x$, $\hat x$, and $\hat \nu$, we choose $\nu$ via the following \emph{interface} function:
	\begin{align}\label{Eq_405a}
		\nu=\nu_{\hat \nu}(x,\hat x,\hat \nu):=K(x-P\hat x)+Q\hat x+\tilde R\hat \nu+L_1\varphi(Fx)-L_2\varphi(FP \hat x),
	\end{align}
	for some matrix $\tilde R$ of appropriate dimension. 
	By employing the equations~\eqref{As_2a},~\eqref{Eq_10a},~\eqref{Eq_14a},~\eqref{Eq_15a} and also the definition of the interface function in~\eqref{Eq_405a}, we simplify
	\begin{align}\notag
		Ax+E\varphi(Fx)+B\nu_{\hat \nu}(x,\hat x, \hat \nu)+Dw -P(\hat A\hat x+\hat E\varphi(\hat F\hat x)+\hat B\hat \nu+\hat D\hat w)
		+(R\varsigma-P\hat R\hat\varsigma),
	\end{align}
	to
	\begin{align}\label{Eq_18a}
		(A&+BK)(x-P\hat x)+Z(Gw-\hat G\hat w)+(B\tilde R-P\hat B)\hat \nu+(BL_1+E)(\varphi(Fx)-\varphi(FP\hat x))+(R\varsigma-P\hat R\hat\varsigma). 
	\end{align}
	From the slope restriction~\eqref{Eq_6a}, one obtains
	\begin{align}\label{Eq_19a}
		\varphi(Fx)-\varphi(FP\hat x)=\delta(Fx-FP\hat x)=\delta F(x-P\hat x),
	\end{align}
	where $\delta$ is a constant and depending on $x$ and $\hat x$ takes values in the interval $[0,b]$. Using~\eqref{Eq_19a}, the expression in~\eqref{Eq_18a} reduces to
	\begin{align}\notag
		((A&+BK)+\delta(BL_1+E)F)(x-P\hat x)+Z(Gw-\hat G\hat w)+(B\tilde R-P\hat B)\hat \nu+(R\varsigma-P\hat R\hat\varsigma). 
	\end{align}
	Using Cauchy- Schwarz inequality,~\eqref{Eq_8a},~\eqref{Eq_12a}, and~\eqref{Eq_13a}, one can obtain the chain of inequalities in~\eqref{Eq_305a} in order to acquire an upper bound. Hence, the proposed $V$ in~\eqref{Eq_7a} is an SStF from  $\widehat \Sigma_{\textsf{nl}}$ to $\Sigma_{\textsf{nl}}$, which completes the proof. 
\end{proof}
\begin{figure*}
	\rule{\textwidth}{0.1pt}
	\begin{align}
	\notag\EE& \Big[V(x(k+1),\hat x(k+1)\,|\,x(k)\!=\!x,\hat x(k)\!=\!\hat x, w(k)\!=\!w, \hat w(k)\!=\!\hat w,\hat \nu(k)\!=\! \hat \nu \Big]-V(x,\hat x)\\\notag 
	&=(x-P\hat x)^T\Big[((A\!+\!BK)\!+\!\delta(BL_1\!+\!E)F)^T\tilde M((A\!+\!BK)\!+\!\delta(BL_1\!+\!E)F)\Big](x-P\hat x)+2 \Big[(x-P\hat x)^T((A\!+\!BK)\\\notag
	&+\delta(BL_1+E)F)^T\Big]\tilde M\Big[Z(Gw-\hat G\hat w)\Big]+2 \Big[(x-P\hat x)^T((A+BK)+\delta(BL_1+E)F)^T\Big]\tilde M\Big[(B\tilde R-P\hat B)\hat \nu\Big]\\\notag
	&+2 \Big[(Gw\!-\!\hat G\hat w)^T Z^T\Big]\tilde M\Big[(B\tilde R\!-\!P\hat B)\hat \nu\Big]+\hat \nu^T(B\tilde R\!-\!P\hat B)^T \tilde M(B\tilde R\!-\!P\hat B)\hat \nu+(Gw\!-\!\hat G\hat w)^T Z^T\tilde M Z(Gw\!-\!\hat G\hat w)\\\notag
	&+\text{Tr}\big(R^T\tilde MR+\hat R^TP^T\tilde MP\hat R\big)-V(x,\hat x)
	=\begin{bmatrix}x-P\hat x\\Gw-\hat G\hat w\\\delta F(x-P\hat x)\\\hat\nu\\\end{bmatrix}^T\\\notag
	&\begin{bmatrix}\notag
	(A\!+\!BK)^T\tilde M(A\!+\!BK) && (A\!+\!BK)^T\tilde MZ && (A\!+\!BK)^T\tilde M(BL_1\!+\!E) && (A\!+\!BK)^T\tilde M(B\tilde R\!-\!P\hat B)\\
	*&& Z^T \tilde MZ && Z^T \tilde M(BL_1\!+\!E) && Z^T \tilde M(B\tilde R\!-\!P\hat B) \\
	*&&*&&(BL_1\!+\!E)^T \tilde M(BL_1\!+\!E)&&(BL_1\!+\!E)^T \tilde M(B\tilde R\!-\!P\hat B) \\
	*&&*&&*&&(B\tilde R\!-\!P\hat B)^T \tilde M(B\tilde R\!-\!P\hat B)
	\end{bmatrix}\\\notag
	&\begin{bmatrix}x-P\hat x\\Gw-\hat G\hat w\\\delta F(x-P\hat x)\\\hat\nu\\\end{bmatrix}+\text{Tr}\Big(R^T\tilde MR+\hat R^TP^T\tilde MP\hat R\Big) -V(x,\hat x)\\\notag
	&\le\notag
	\begin{bmatrix}x-P\hat x\\Gw-\hat G\hat w\\\delta F(x-P\hat x)\\\hat\nu\\\end{bmatrix}^T\begin{bmatrix}
	\widehat\kappa\tilde M+C_2^T\bar X^{22}C_2 & C_2^T\bar X^{21} & -F^T & 0\\
	\bar X^{12}C_2 & \bar X^{11} & 0 & 0\\
	-F & 0 & \frac{2}{b} & 0 \\ 0 & 0& 0& \tilde k(B\tilde R-P\hat B)^T \tilde M(B\tilde R-P\hat B) \notag
	\end{bmatrix}\begin{bmatrix}x-P\hat x\\Gw-\hat G\hat w\\\delta F(x-P\hat x)\\\hat\nu\end{bmatrix}\\\notag
	&+\text{Tr}\Big(R^T\tilde MR+\hat R^TP^T\tilde MP\hat R\Big)-V(x,\hat x)=-(1-\widehat\kappa) (V(x,\hat x))-2\delta(1-\frac{\delta}{b})(x-P\hat x)^TF^TF(x-P\hat x)\\\notag
	&+\tilde k\Vert\sqrt{\tilde M}(B\tilde R-P\hat B)\nu\Vert^2+\begin{bmatrix}Gw-\hat G\hat w\\C_2x-H\hat C_2\hat x\end{bmatrix}^T\begin{bmatrix}
	\bar X^{11}&\bar X^{12}\\
	\bar X^{21}&\bar X^{22}
	\end{bmatrix}\begin{bmatrix}Gw-\hat G\hat w\\C_2x-H\hat C_2\hat x\end{bmatrix}+\text{Tr}\Big(R^T\tilde MR+\hat R^TP^T\tilde MP\hat R\Big)\\\notag
	&\le -(1-\widehat\kappa) (V(x,\hat x))+\tilde k\Vert\sqrt{\tilde M}(B\tilde R-P\hat B)\Vert^2\Vert\hat \nu\Vert^2
	+\begin{bmatrix}Gw-\hat G\hat w\\C_2x-H\hat C_2\hat x\end{bmatrix}^T\begin{bmatrix}
	\bar X^{11}&\bar X^{12}\\
	\bar X^{21}&\bar X^{22}
	\end{bmatrix}\begin{bmatrix}Gw-\hat G\hat w\\C_2x-H\hat C_2\hat x\end{bmatrix}\\\label{Eq_305a}
	&+\text{Tr}\Big(R^T\tilde MR+\hat R^TP^T\tilde MP\hat R\Big).
	\end{align}
	\rule{\textwidth}{0.1pt}
\end{figure*}
\begin{proof}\textbf{(Lemma~\ref{lem:ineq_DFA})}
	Suppose $\hat y(\cdot)\vDash_{\Lab^\epsilon} {\hat\phi}$ over time interval $[0,T_d]$. According to the construction of DFA $\mathcal{A}_{\hat\phi}$, $q_{abs}$ is an absorbing state and not an accepting state, thus $\Lab^\epsilon (\hat y(k))\neq \phi_\circ$,  $\forall k\in[0,T_d]$. 
	Then $\Lab^\epsilon (\hat y(k)) \in \Sigma_{\textsf{a}}$, $\forall k\in[0,T_d]$.
	Assume $\Lab^\epsilon (\hat y(k))=a$ then $\hat y(k) \in [\Lab^{-1}(a)]^\epsilon$.
	Since we know that $$\sup_{0\le k\le T_d}\Vert y(k)-\hat y(k) \Vert < \epsilon,$$ then according to the definition of \mbox{$\epsilon$-perturbed} sets $y(k) \in \Lab^{-1}(a)$ which gives $\Lab(y(k))=a$.
	Thus $\Lab(y(\cdot)) = \Lab^\epsilon(\hat y(\cdot))$ and having $\hat y(\cdot)\vDash_{\Lab^\epsilon} {\hat\phi}$ guarantees $y(\cdot)\vDash_{\Lab} \phi$ due to the particular construction of $\hat\phi$.

\end{proof}
\begin{proof}\textbf{(Theorem~\ref{Thm: Lowerbound})}	
	According to Lemma~\ref{lem:ineq_DFA}, $y(\cdot)\nvDash_{\Lab} \phi$ results in $\hat y(\cdot)\nvDash_{\Lab^\epsilon}\hat\phi$ over time interval $[0,T_d]$ or $$\sup_{0\le k\le T_d}\Vert y(k)-\hat y(k) \Vert \geq \epsilon.$$ Then
	\begin{align*}
		&\PP (y(\cdot)\nvDash_{\Lab} \phi) \leq \PP(\hat y(\cdot)\nvDash_{\Lab^\epsilon} \hat\phi) + \overbrace{\PP (\sup_{0\le k\le T_d}\Vert y(k)-\hat y(k) \Vert \geq \epsilon)}^{\le\delta} \\
		&\Rightarrow 1-\PP (y(\cdot)\vDash_{\Lab} \phi) \leq 1-\PP(\hat y(\cdot)\vDash_{\Lab^\epsilon} \hat\phi)+\delta\\
		&\Rightarrow \PP(\hat y(\cdot)\vDash_{\Lab^\epsilon} \hat\phi)-\delta \leq \PP (y(\cdot)\vDash_{\Lab} \phi), 
	\end{align*}
	which completes the proof.
\end{proof}

\end{document}